\documentclass[a4paper,10pt]{article}
\usepackage[margin=1in]{geometry}
\usepackage{amsmath,amssymb,amsthm}
\usepackage{graphicx} 
\usepackage{braket}
\usepackage{dsfont}
\usepackage[normalem]{ulem}
\usepackage[allcolors=blue,colorlinks=true,hyperindex=true]{hyperref}
\usepackage{tabularx}
\usepackage{multirow}
\usepackage{array}
\usepackage{nicematrix}
\usepackage{subcaption}
\usepackage{booktabs}
\usepackage{siunitx}
\usepackage{graphicx}    
\usepackage{subcaption}  
\usepackage[ruled, lined, linesnumbered, commentsnumbered, longend]{algorithm2e}
\usepackage{pgfplots}
\usepackage{subcaption}
\usepackage{authblk}

\usepackage[utf8]{inputenc}
\usepackage{tcolorbox}
\tcbuselibrary{listings, breakable, skins}
\usepackage{xcolor}

\usepackage{tcolorbox}
\tcbuselibrary{listings, breakable}
\usepackage{xcolor}
\usepackage{inconsolata} 

\definecolor{ghbackground}{HTML}{F6F8FA}
\definecolor{ghkeyword}{HTML}{D73A49}
\definecolor{ghcomment}{HTML}{6A737D}
\definecolor{ghstring}{HTML}{032F62}
\definecolor{ghidentifier}{HTML}{005CC5}

\lstdefinestyle{githubstyle}{
  language=Python,
  basicstyle=\ttfamily\small,
  keywordstyle=\color{ghkeyword}\bfseries,
  commentstyle=\color{ghcomment}\itshape,
  stringstyle=\color{ghstring},
  identifierstyle=\color{ghidentifier},
  showstringspaces=false,
  breaklines=true,
  tabsize=4,
  frame=none,
  numbers=none
}

\newtcblisting{codeblock}{
  listing only,
  breakable,
  enhanced,
  colback=ghbackground,
  colframe=black!10,
  boxrule=0pt,
  arc=4pt,
  outer arc=4pt,
  top=4pt,
  bottom=4pt,
  left=6pt,
  right=6pt,
  boxsep=2pt,
  listing options={style=githubstyle},
  fonttitle=\bfseries,
  attach title to upper,
  before upper={\captionsetup{type=listing}},
}

\usepackage{caption}

\usepackage[capitalise]{cleveref}

\usepackage{float}

\theoremstyle{definition}

\newtheorem*{hypo}{Hypothesis}
\newtheorem*{remark}{Remark}

\newcommand{\nmax}{N_{\mathrm{max}}}

\title{A Practical Guide to using Pauli Path Simulators for Utility-Scale Quantum Experiments}

\author[1]{Hrant Gharibyan}
\author[1,2,3]{Siddharth Hariprakash}
\author[1]{Mohammed Zuhair Mullath}
\author[1]{Vincent P. Su}

\affil[1]{BlueQubit Inc, San Francisco, CA 94105, USA}
\affil[2]{Leinweber Institute for Theoretical Physics and Department of Physics, University of California, Berkeley, California 94720, USA}
\affil[3]{Physics Division, Lawrence Berkeley National Laboratory, Berkeley, California 94720, USA}

\begin{document}

\maketitle
\begin{abstract}
In this this paper we present an inexpensive protocol to perform runtime and memory estimation for large-scale experiments with Pauli Path simulators (PPS). Additionally, we propose a conceptually simple solution for studying whether PPS can be used as a scientific discovery tool, rather than reproducing existing answers.

We start by analyzing the dynamics of the Pauli coefficients tracked in the Heisenberg picture. In addition to surprisingly generic convergence features of the coefficient distributions, we find certain regularities that allow for extrapolation of memory and runtime requirements for smaller and smaller coefficient truncation parameter $\delta$.

We then introduce a framework for understanding convergence in the absence of rigorous error guarantees on PPS. Combined with runtime analysis, we propose bifurcating quantum simulation problems broadly into two classes, based on whether there is apparent convergence of expectation values as a function of $\delta$. This serves as a way for practitioners to understand where their problem falls on the frontier of classical simulability. In the case without apparent convergence, PPS may still serve useful as a Monte Carlo-like estimate. Applied to IBM's utility-scale experiments, we show parameter regimes where both behaviors are realized. Some of our key findings challenge conventional intuition: reducing $\delta$ does not always improve accuracy, and deeper quantum circuits may actually be easier to simulate than shallower ones.

Our analysis provides the first systematic approach to assess PPS reliability \textit{without} assuming a groundtruth quantum answer. The BlueQubit SDK implementing these methods has been released publicly, offering researchers a comprehensive toolkit for evaluating this frontier classical simulation approach. These results establish practical guidelines for when PPS can serve as a reliable verification tool versus when it should be used as a complementary estimate alongside quantum experiments.

  \end{abstract}
\tableofcontents

\section{Introduction}

Simulating quantum systems remains one of the central challenges in computational physics, computational chemistry, and quantum information science. Despite the exponential growth of Hilbert space with system size, classical simulation methods continue to provide essential insights into quantum dynamics, verification of quantum hardware, and development of novel quantum algorithms. While tensor network methods \cite{Vidal:2003pmm,Haegeman:2011zz,Orus2014117,Haegeman:2016gfj,biamonte2017tensornetworksnutshell} and their variants \cite{Tindall:2023cqi, Tindall_2023} have long dominated this space -- particularly for low-entanglement systems -- recent innovations have begun to chart alternative directions that can tackle problems outside the reach of conventional approaches.

One such innovation is the Pauli Path Simulation (PPS) framework,\footnote{Also commonly referred to as Pauli propagation or Sparse Pauli Dynamics (SPD) framework.} which has emerged as a powerful method for simulating quantum systems in the Heisenberg picture. Instead of evolving quantum states forward in time -- as done in traditional statevector based simulations -- PPS propagates observables backward through a quantum circuit. The observable is decomposed into a sum of Pauli strings, and its evolution is approximated by sampling and propagating the most significant terms along discrete Pauli paths determined by the circuit unitaries. While a full expansion involves up to $4^n$ Pauli terms for an $n$-qubit system -- compared to $2^n$ amplitudes in the Schrödinger picture -- this exponential overhead can be mitigated through judicious truncation strategies that preserve accuracy while greatly reducing computational cost.

Originally inspired by the NISQ era quantum computers -- where quantum circuits are often shallow and dominated by noise -- it is natural to consider the {\it size}, or Hamming  weight, of Pauli operators when deciding which terms to truncate. One of the simplest truncation schemes is to discard operators that act non-trivially on more than $k$ qubits. For instance, \cite{PhysRevLett.133.120603,Fontana:2023wvl,Schuster:2024jds, Aharonov_2023, angrisani2025} show that the error introduced by this approximation decays exponentially with increasing $k$. This result was later extended to an ensemble of noiseless circuits as well (see, e.g., \cite{Angrisani:2024nrg, lerch2024}). Since its initial conception, additional approximation strategies have been explored, such as {\it coefficient-based truncation} \cite{Begusic:2023thb,PRXQuantum.6.020302}, where Pauli terms with amplitudes below a specified threshold, $\delta$, are systematically discarded. More recently, Rudolph et al. \cite{Rudolph:2025gyq} presented a comprehensive account of Pauli propagation methods, including their mathematical foundations, truncation strategies, and practical implementation as a Julia package.

One of the first notable experimental applications of Pauli Path simulations was performed in \cite{Begusic:2023jwi} to reproduce the expectation values in a 127 qubit dynamical quantum kicked Ising model, originally obtained using IBM hardware \cite{Kim:2023bwr}. While statevector simulation is limited to a few dozens of qubits, even other leading approximations such as generic tensor network-based matrix product states failed to correctly capture the entanglement at later stages of the time evolution, hinting at a potential demonstration of classical intractability.\footnote{Only highly specialized tensor network approaches \cite{Tindall:2023cqi} tailored to IBM’s specific heavy-hexagon connectivity yielded approximate agreement.} However, as is typical, new quantum results lead to innovation on the classical algorithm side as well, prompting others to reproduce the results with very moderate resources.

PPS introduces a critical question that has received limited systematic attention: When can we trust the results? Unlike exact simulation methods, PPS relies on approximations whose errors are difficult to bound a priori. A smaller truncation threshold $\delta$ should in principle yield more accurate results, yet our investigations reveal this intuition to be misleading. Similarly, one might expect deeper quantum circuits to be universally harder to simulate, but we find examples where the opposite holds true. These counterintuitive behaviors point to a fundamental gap in our understanding of when and how PPS can reliably estimate quantum observables.

Our primary goal of this paper is to provide a practical framework that helps quantum researchers evaluate the utility of PPS for their problem. In addition to serving as a pedagogical text for an overview of this method, we present crucial memory and runtime analysis that will help users better understand the computational resources required to deploy this method. An often overlooked aspect of PPS is the fluctuation of expectation values as a function of the truncation parameter. Combined with a lack of practical error bounds, this can make PPS less reliable as a discovery tool, but still valuable as a complement to quantum hardware runs. Our framework involves studying this convergence with extremely modest CPU resources to diagnose whether PPS has \textit{apparently} converged and also to extrapolate the computational demands of finer-resolution simulations.
We also introduce the newly integrated BlueQubit SDK for Python, which enables scalable PPS simulations with minimal infrastructure requirements. As a practical demonstration, we include a working code snippet that applies our convergence framework to IBM’s utility circuit in App.~\ref{app:code}.

For researchers considering classical simulation approaches, a natural question arises: when should one choose PPS over tensor network methods? Currently, neither approach strictly dominates the other across all simulation scenarios. The methods exhibit complementary strengths depending on the specific computational task. When evaluating multiple observables for a given circuit, PPS requires separate runs for each observable, while tensor network methods typically compute a single approximate time-evolved state from which multiple expectation values can be efficiently extracted. However, this advantage comes with trade-offs in complexity: for the most challenging simulation problems, tensor networks often require extensive tuning of bond dimensions, contraction schemes, and network architectures, whereas Pauli propagation offers more straightforward control through relatively simple truncation parameters.

\textbf{Paper Organization}: The remainder of this paper is structured as follows. Section 2 presents an accessible overview of the PPS technique, emphasizing the branching mechanism that drives computational complexity. Section 3 develops our theoretical framework for predicting resource requirements, revealing the general power-law structure of coefficient distributions and demonstrating how this enables accurate extrapolation of memory and runtime costs from brief, inexpensive experiments. Section 4 introduces our convergence assessment protocol and applies it systematically to IBM's kicked Ising circuits, revealing the two distinct regimes where PPS either apparently converges or provides useful Monte Carlo-like estimates. Throughout, we emphasize the counterintuitive findings that challenge conventional assumptions about classical simulation. Section 5 discusses the broader implications for quantum computing and the role of PPS in distinguishing between classically tractable and genuinely quantum-advantaged problems.

\section{The Branching and Merging Dynamics of PPS}

In this section, we provide an overview of the PPS simulation technique used to generate the numerical results presented in this work.\footnote{For additional description of the technique, see for example \cite{Begusic:2023thb,PRXQuantum.6.020302,Rudolph:2025gyq}.} At the highest level, this is simply working backwards through a circuit, applying the Heisenberg picture to an observable one gate at a time. This step-by-step procedure can be thought of as a flow from Pauli terms with large coefficients to many more Pauli terms with smaller coefficients. This intuitive picture will be made more concrete numerically in~\cref{sec:proliferation}.

In the standard Schrödinger picture of quantum mechanics, states evolve while observables remain fixed. For a quantum circuit applied to an initial state $|\psi_0\rangle$. In this work, we set $|\psi_0\rangle = \ket{0}$ and hence computing the expectation value of an observable $O$ requires:
\begin{align}
    \langle O \rangle = \bra{0} U^\dagger O U \ket{0},
\end{align}
where $U$ represents the unitary transformation of the entire circuit. Most common quantum simulation approaches track evolution of the quantum state storing an exponentially large statevector. The PPS method instead employs the Heisenberg picture, where observables evolve backward through the circuit while states remain fixed:
\begin{align}
\langle O \rangle = \bra{0} U^\dagger O U \ket{0} = \bra{0} O(t) \ket{0},
\end{align}
where $O(t) = U^\dagger O U$ represents the time-evolved observable. This transformation allows us to calculate expectation values without explicitly storing the full quantum state.

Without loss of generality, the $n$-qubit unitary $U$ representing the unitary transformation of a given circuit can be represented as a series of Pauli rotations as follows:
\begin{align}
    \label{eq:circuit_unitary}
    U = \prod_{j\in[J]}U_j(\theta_{j}) = \prod_{j\in[J]}\mathrm{e}^{-i\theta_j \sigma_j/2},
\end{align}
where each $\sigma_j$ is an $n$-qubit Pauli operator, and each $\theta_j \in [-\pi/4,\pi/4]$. The restriction on the angles $\theta_j$ can be accomplished using the method of Clifford recompilation as described for example in \cite{Qassim2019clifford,Begusic:2023jwi}.

Let $O_k$ be the observable after the action of $k$ gates:
\begin{align}
    \label{eq:O_m}
    O_k := U_k^{\dagger}U_{k-1}^{\dagger}\dots U_1OU_1U_2 \dots U_k.
\end{align}
Using the Pauli basis, $O_{k}$ can always be decomposed as
\begin{align}
    \label{eq:pauli_decomp_k}
    O_k = \sum_{P\in\mathcal{P}_k}c_{P}^{(k)}P,
\end{align}
where $\mathcal{P}_k$ is a subset of all possible $n$-qubit Pauli operators with non-zero coefficients.

The key to understanding PPS behavior lies in how Pauli operators interact with circuit gates. We partition the Pauli terms based on their commutativity with the next gate generator $\sigma_{k+1}$:
\begin{align}
    \mathcal{P}^{comm}_{k} &:= \{P \in\mathcal{P}_k\hspace{1mm} \lvert \hspace{1mm} [P, \sigma_{k+1}]=0\}, \\
    \mathcal{P}^{anti}_{k} &:= \{P \in\mathcal{P}_k\hspace{1mm} \lvert \hspace{1mm} \{P, \sigma_{k+1}\}=0\}.
\end{align}
This partition determines the computational cost: commuting terms pass through unchanged, while anti-commuting terms branch, possibly creating new Pauli strings. Specifically:
\begin{align}
\label{eq:U_k+1}
U_{k+1}^{\dagger} P U_{k+1} =
\begin{cases}
P &\quad \text{if} \quad P \in \mathcal{P}^{comm}_{k} \\
\cos(\theta_{k+1})P + \sin(\theta_{k+1})P' &\quad \text{if} \quad P\in \mathcal{P}^{anti}_{k}.
\end{cases}
\end{align}
Here, $P' = i\sigma_{k+1} P$ is the conjugated Pauli string. This branching is the fundamental source of exponential growth in PPS simulations—each anti-commuting term potentially doubles the number of tracked operators.
Thus, the size of the relevant set of Paulis grows as $\mathcal{P}_{k+1} = \mathcal{P}_{k} \cup \sigma_{k+1} \mathcal{P}^{anti}_{k}$.

The coefficient evolution follows a branching-and-merging pattern. For a Pauli term $P$, its updated coefficient $c_{P}^{(k+1)}$ depends on multiple factors: whether it existed before the gate, whether it commutes with the gate, and whether other terms branch into it:
Using that $\sigma_{k+1}^2 = 1$, this merging can be pre-determined by checking whether $P' = i\sigma_{k+1}P \in \mathcal{P}^{anti}_{k} $. 

More explicitly,
\begin{align}
    \label{eq:branching-and-merging} 
    c_{P}^{(k+1)} =
    \begin{cases}
    \begin{alignedat}{5}
        &c_{P}^{(k)} &&                  &&                 &&\quad \text{if } P \in \mathcal{P}^{comm}_{k} 
        \\
        &c_{P}^{(k)} \cos(\theta_{k+1}) &&                  &&                 &&\quad \text{if } P \in \mathcal{P}^{anti}_{k} &&\text{ and } P' \notin \mathcal{P}_{k} \\
        &c_{P}^{(k)} \cos(\theta_{k+1}) &&\,+\,            && c_{P'}^{(k)} \sin(\theta_{k+1}) &&\quad \text{if } P \in \mathcal{P}^{anti}_{k} &&\text{ and } P' \in \mathcal{P}_{k} \\
        &                               &&                  && c_{P'}^{(k)} \sin(\theta_{k+1}) &&\quad \text{if } P \notin \mathcal{P}_{k} &&\text{ and } P' \in \mathcal{P}^{anti} _{k}
    \end{alignedat}
    \end{cases}
\end{align}

In this manner, the Pauli decomposition of the evolved observable can be iteratively updated by traversing through (in order) and evolving the observable by the unitaries $U_j$. Generically, this will lead to a proliferation of terms stemming from the anti-commuting set. However, in the case of Clifford operations, the angles will be restricted to special values that prevent this branching. Hence, the presence of a large number of non-Clifford operations can lead to rapid growth in the total number of terms present in the Pauli decomposition of the evolved observable.
For details on representing the Pauli strings with bit arrays, see Appendix~\ref{app:implementation}.

Up to this point, the simulation technique is exact as every possible term arising throughout the evolution has been accounted for. The approximation made by the PPS method is to impose a $\delta \in \mathbb{R}^+$ truncation on the norms of the coefficients in the Pauli basis of the evolved observable at any given step in the evolution. Thus, the coefficients $c_{P}^{(k)}$ appearing in \cref{eq:pauli_decomp_k} are now required to satisfy $|c_{P}^{(k)}| \geq \delta$. Upon evolving the observable $O_k$ by the unitary $U_{k+1}$, we truncate the Pauli decomposition shown in \cref{eq:U_k+1} by imposing the condition $|c_{P}^{(k+1)}| \geq \delta$. In this manner, one can control the (in general) exponential growth in the number of terms appearing in the Pauli decomposition of the evolved observable.

Though straightforward from~\cref{eq:branching-and-merging}, we note the fact that the coefficients are always decreasing in magnitude based on these dynamics. Because it will be useful in the following section, consider treating the sum~\cref{eq:pauli_decomp_k} as a continuous density over coefficients. A single gate application only has a discrete set of possibilities for affecting a given coefficient, and hence for its impact on the density. In the limit where every angle in the circuit is the same, the density will be a series of Dirac $\delta$ spikes. In the opposite limit, if the angles are generically uncorrelated, then we find that this density can actually be well-modeled by a power law. It turns out the latter is a more generic behavior.

\section{Modeling the Proliferation of Pauli Operators}\label{sec:proliferation}

This section presents our framework for understanding and predicting the computational resource requirements of Pauli Path Simulators with coefficient truncation. The core challenge we address is determining whether a quantum circuit can be simulated within available memory and time constraints before committing to expensive computational runs. Our analysis reveals that the seemingly complex dynamics of Pauli string proliferation can be modeled by surprisingly simple mathematical patterns. 

In particular, we show a simple observation regarding the distribution obeyed by the coefficients of the evolved observable in the Pauli basis enables accurate predictions of memory and runtime requirements for a class of circuits. For the researcher interested in applying PPS to their quantum circuit calculation, we translate these theoretical insights into a practical algorithm for predicting resource needs using only brief test runs. 

\textit{A word on notation:} For an observable $O$ with Pauli-sum representation $\sum_{P} c_P P$ in the \textit{unnormalized} $n$-qubit Pauli basis, we shall be writing $\|O\|$ to denote the ``raw norm" of $O$ given by $(\sum_P |c_P|^2)^{1/2}$. It is related to the standard Hilbert-Schmidt norm by $\|O\|_2 = (\mathrm{Tr}(O^\dagger O))^{1/2} = 2^{n} \|O\|$.

\subsection{Evolution of the distribution of Pauli coefficients}

\begin{figure}
    \centering
    \includegraphics[width=0.95\linewidth]{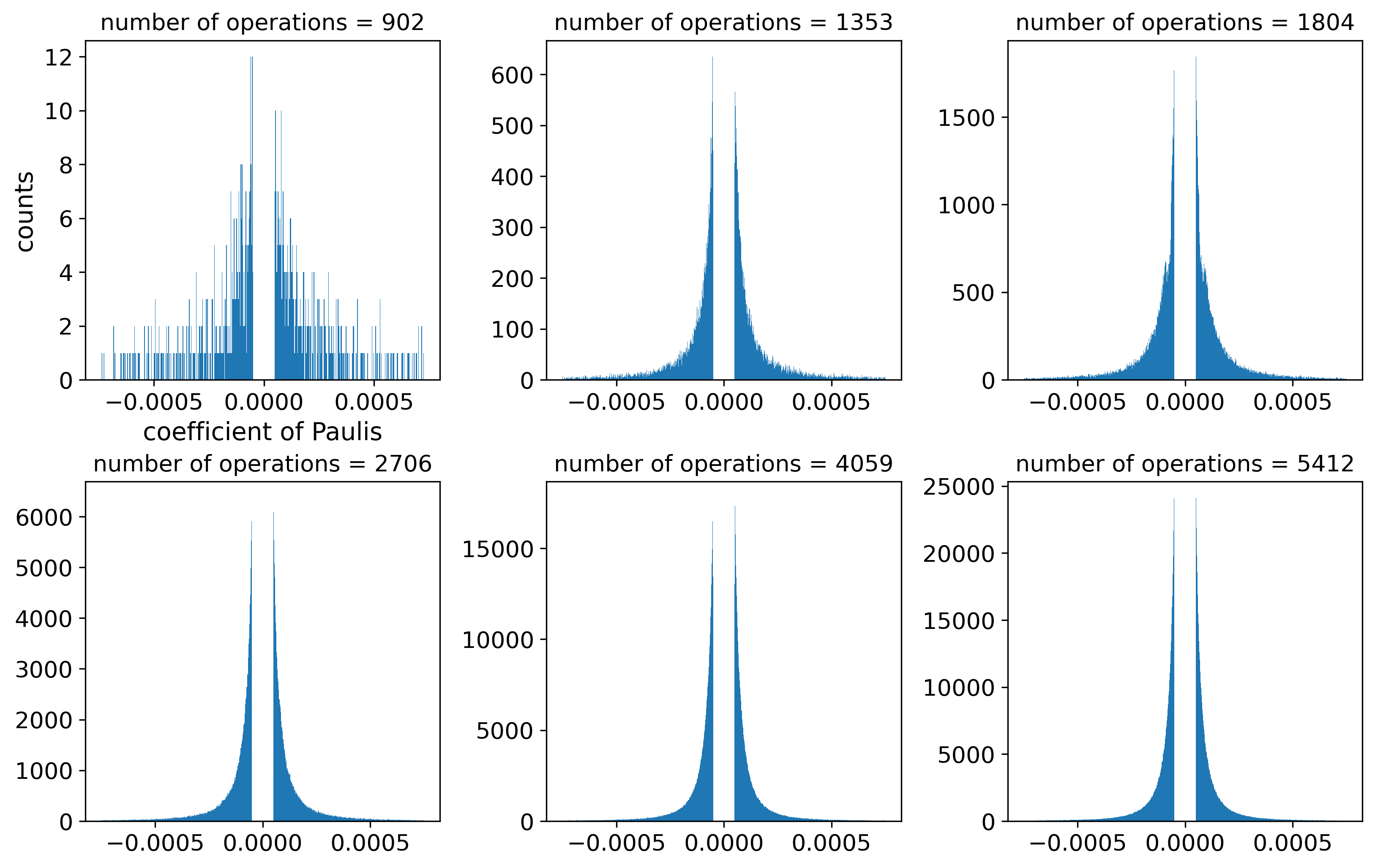}
    \caption{Evolution of the distribution of Pauli coefficients shown at different stages in the execution of a circuit of the form \cref{eq:kicked-ising} with each $\theta_{X}$ chosen by sampling uniformly from $[-\pi/4,\pi/4]$, $\theta_{ZZ} = -\pi/2$, $T = 20$, and a coefficient threshold $\delta = 5\times10^{-5}$. We set the initial observable to be the single weight operator $\langle Z_{62} \rangle$ and observe that upon evolving using sufficiently many gate operations, the distribution appears to possess some invariant structure.}
    \label{fig:distribution-invariance}
\end{figure}

\begin{figure}
    \centering
    \includegraphics[width=1.0\linewidth]{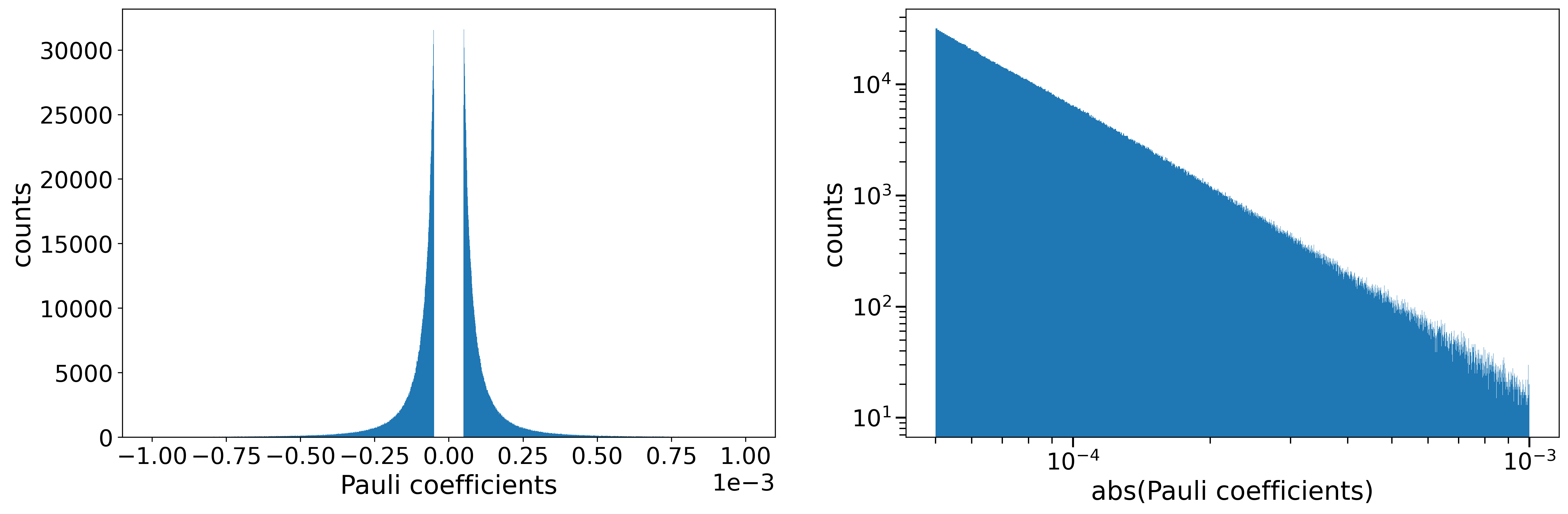}
    \caption{(left) Distribution of Pauli Coefficients corresponding to the set up described in \cref{fig:distribution-invariance} and (right) the distribution of the absolute values of these coefficients. The data shown was captured after the execution of 5412 out of the 5420 total gates in the circuit. There are approximately $2.5$ million unique Paulis present in the evolved observable and we use 2048 bins to plot the histograms. The distribution shown on the right can be approximated by a power law with larger deviations towards the tail.}
    \label{fig:coeff-distribution}
\end{figure}

We start by considering the kicked Ising dynamics studied by IBM in \cite{Kim:2023bwr} on quantum hardware as well as in \cite{Tindall:2023cqi,Liao:2023onw,Patra:2023qoo} using state of the art tensor network methods. This Hamiltonian simulation problem amounts to implementing unitaries of the form
\begin{align}\label{eq:kicked-ising}
    U = \prod_{t\in[T]}\left(\prod_{\langle i,j \rangle}e^{-\mathrm{i}\theta_{ZZ}Z_iZ_j/2}\prod_{i}e^{-\mathrm{i}\theta_{X}X_i/2}\right),
\end{align}
where $T$ denotes the total number of Trotter steps in the evolution and $\langle i,j \rangle$ indicates nearest neighbors on IBM's 127 qubit heavy hex lattice. In existing literature, $\theta_{X}$ is generally fixed to be a constant throughout a single circuit instance (i.e. the single qubit rotation gates are chosen to be correlated). However, to break translation invariance and give a slightly more generic circuit, in this section we consider the case where we choose each of the single qubit rotation angles $\theta_X$ by sampling uniformly from $[-\pi/4,\pi/4]$\footnote{\cref{app:correlated_distributions,app:analytical} provide numerical evidence suggesting that our findings in this section can be extended to the correlated case as well with some caveats as explained in \cref{app:large_deviations}}. This randomly generated case more closely resembles a broader class of brickwork variational circuits. As in previous work, we set $\theta_{ZZ} = -\pi/2$, i.e. the two-qubit rotations are chosen to be correlated and Clifford. 

Throughout this work, we will set the initial observable $O = Z_{62}$. For a depth $T=20$ circuit we evolve $O$ using the unitary shown in \cref{eq:kicked-ising} and plot the distribution of Pauli coefficients of the evolved observable at different stages in the circuit evolution in \cref{fig:distribution-invariance}. From \cref{eq:branching-and-merging} we expect that the circuit operations will make the Pauli coefficients of the evolved observable smaller (in magnitude), thus explaining the increasing concentration at smaller (in absolute value) coefficient values as we traverse the circuit. About a third of the way through the circuit, we observe that the shape of the distribution looks roughly invariant. Focusing on a point in the evolution after which the shape of this distribution of Pauli coefficients appears invariant, we see that the absolute values of these Pauli coefficients (see \cref{fig:coeff-distribution}) appear to be well approximated by a truncated power law of the form
\begin{equation}\rho(t) = \frac{A}{|t|^{m+1}}, \; |t| > \delta, \label{eq: power_law} \end{equation}
where $A$ is a normalization constant, given by $A=m\delta^m/2$. In \cref{app:analytical} we demonstrate similar behavior if we instead set $\theta_X = \pi/4$ i.e. maximally non Clifford and show that upon making mild assumptions about the similarity of the sets $\mathcal{P}^{\text{comm}}_{k}$ and $\mathcal{P}^{\text{anti}}_{k}$ one can show that such a power law density yields a steady state solution to the dynamics specified by \cref{eq:branching-and-merging}.

\subsection{Bounding the growth of Pauli terms}\label{subsec:power_law_bounds}

\begin{figure}
    \centering
    \begin{subfigure}{0.475\textwidth}
        \centering
        \includegraphics[width=\linewidth]{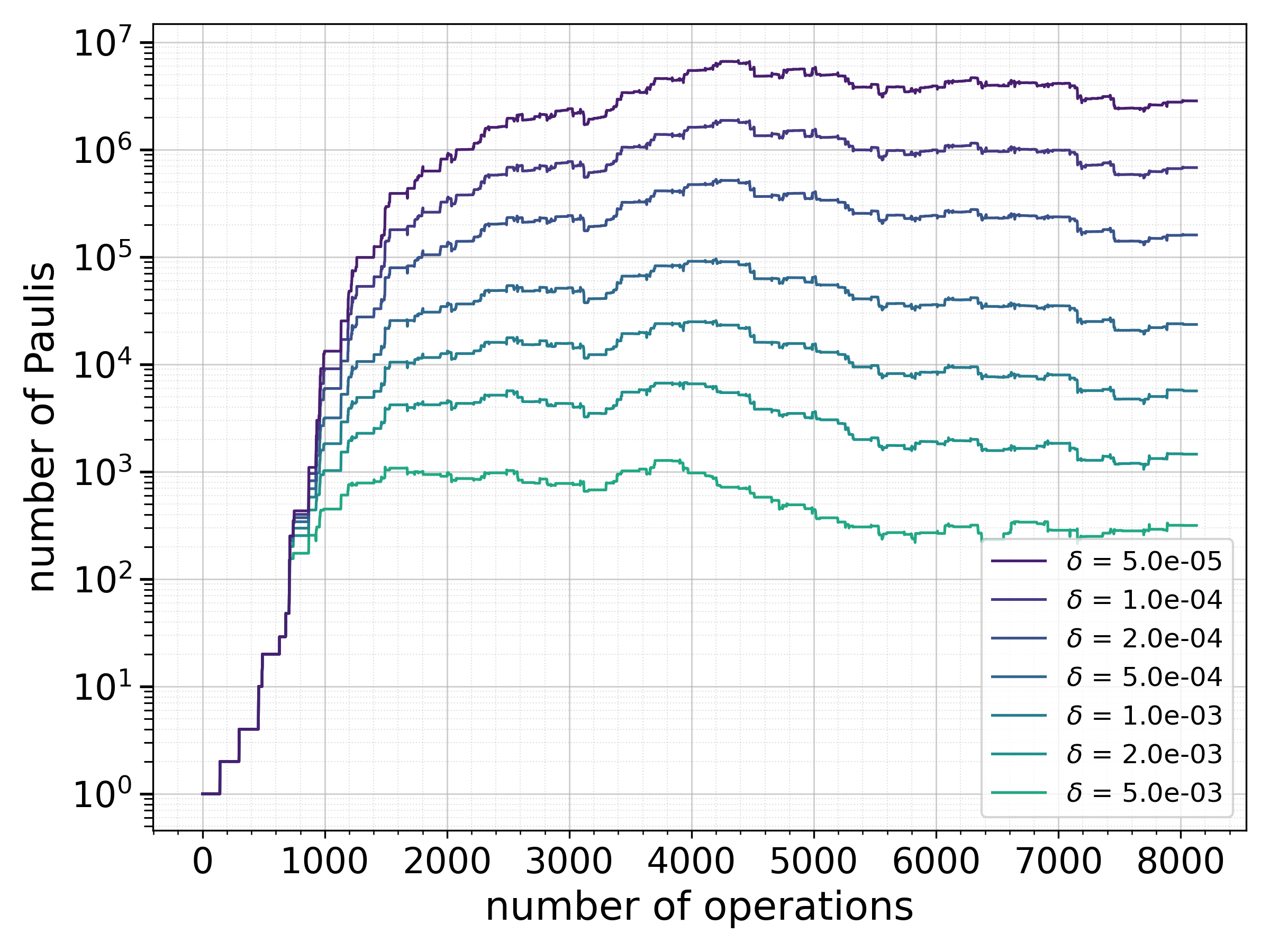}
        \label{fig:growth of Paulis random rx}
    \end{subfigure}
    \begin{subfigure}{0.475\textwidth}
        \centering
        \includegraphics[width=\linewidth]{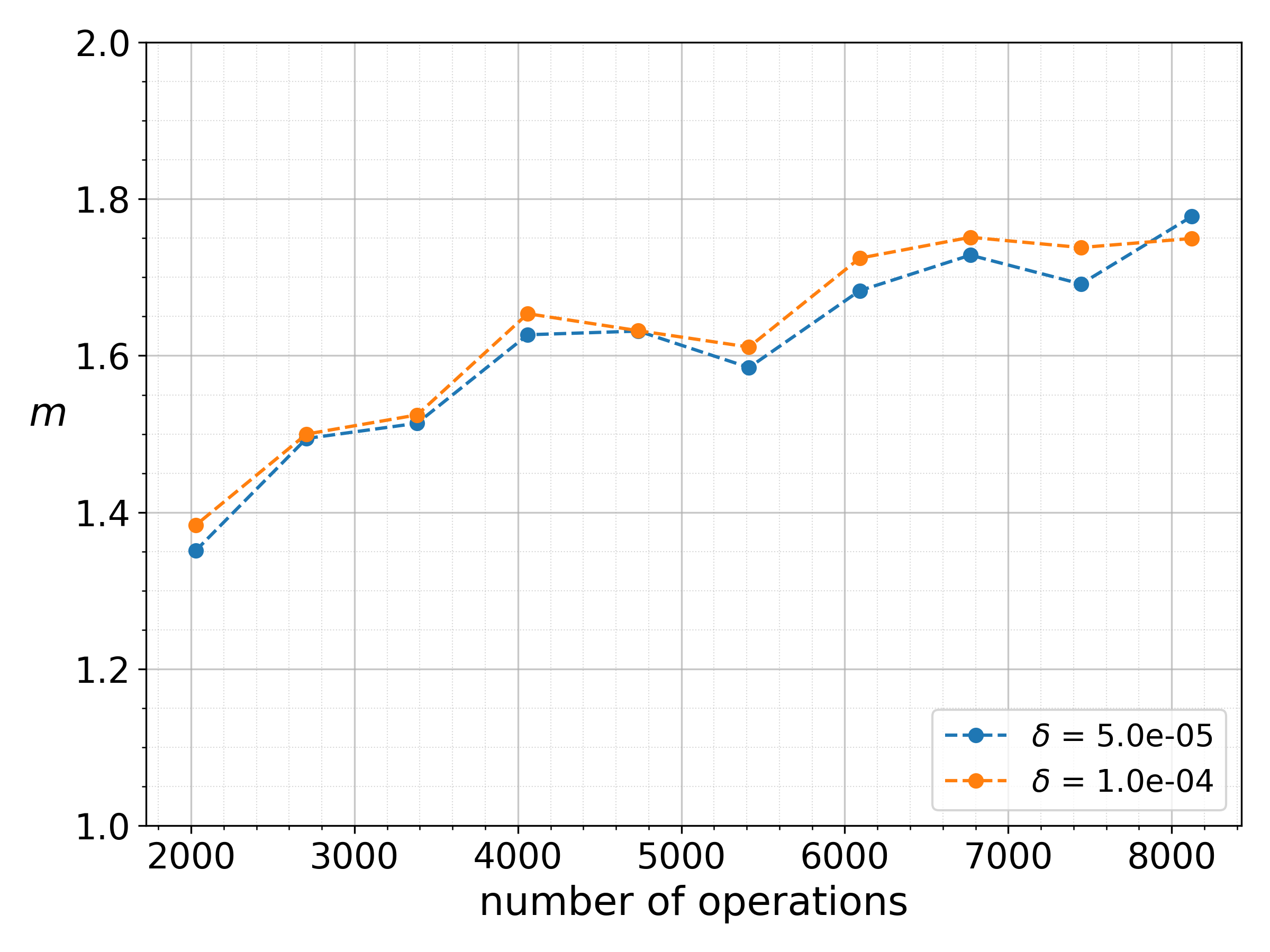}
    \end{subfigure}
    \caption{Tracking the growth of Pauli's and the shape of the power-law coefficient distribution. Left: growth of number of Pauli terms in the simulation with $\theta_X$ sampled uniformly from $[\pi/4,\pi/4]$, $\delta = 5\times 10^{-5}$, $T = 30$; right: power-law exponent $m$ varies over the execution of the same circuit. The values of $m$ are estimated directly from the coefficient distribution of Pauli terms at various points in the circuit execution. For more details about this fit, see \cref{app:m_estimation_challenges}.}
    \label{fig:pauli-growth-and-varying-m}
\end{figure}

In this section, we use the specific form of the power law shown in \cref{eq: power_law} to understand the growth of the number of unique Paulis present in the evolved observable as a function of the number of gate operations used to perform the evolution. In particular, \cref{fig:pauli-growth-and-varying-m} shows examples of such \emph{Pauli growth curves} along with the variation in the power law parameter $m$ as a function of the number of gate operations. There are two interesting observations we make from this set of plots.
\begin{enumerate}
    \item The shapes of the Pauli growth curves (after sufficiently many gate applications) appear roughly similar with vertical shifts that look regular with variations in $\delta$. We also note that the maximum of each curve throughout the evolution, denoted by $\nmax$, also visually appears to be regularly spaced on the vertical axis as a function of $\delta$. \cref{fig: growth of paulis} shows more examples of this observations using different experimental setups.
    \item The parameter $m$ that determines the exact form of the power law \cref{eq: power_law}, as shown on the right, exhibits variations as a function of both $\delta$ and the number of gate applications.
\end{enumerate}
In this section, we provide one possible explanation for the first observation. In \cref{app:analytical} we identify key parameters that explain why the growth curves, although similar, show minor deviations from each other. For more on the second observation as well as subtleties that arise when estimating $m$ see \cref{app:m_estimation_challenges}. 

We begin by observing that imposing the constraint that all surviving terms post truncation have magnitude greater than $\delta$, the following trivial bound holds: 
\begin{equation}
    \nmax \leq \frac{||O||^2}{\delta^2}
\label{eq: trivial-bound} 
\end{equation}
To improve on this analysis, we observe that the average value of the sum computing the norm $\|O_k\|^2$ (where $O_k$ is the evolved observable given by \cref{eq:O_m,eq:pauli_decomp_k}) is a Riemann sum approximation of the second moment of the distribution of the absolute values of the Pauli coefficients $\rho$: 
\begin{equation} 
2\int_\delta ^{\|O \| } t^2 \rho(t) dt \approx \frac{1}{N_k} \sum_{P \in \mathcal P_k} |c_P^{(k)}|^2 = \frac{ \|O_k \|^2}{N_k}. \label{eq: second moment estimate} 
\end{equation}
Using the general definition of $\rho$, given by \cref{eq: power_law}, we thus have that (note the $k$ dependence on $m$ has been assumed due to the second observation made based on \cref{fig:pauli-growth-and-varying-m}):
\begin{align}
    m_k&\delta^{m_k} \int_\delta^{\|O \|} t^2 \frac{dt}{t^{m_k+1}} \approx \frac{ \| O_k \|^2}{N_k} \\
    \label{eq:N_k}
    \implies N_k &\approx \frac{2-m_k}{m_k}\,  \frac{\| O_k \|^2}{\delta^{m_k}(\| O \|^{2-m_k} -\delta^{2-m_k})},
\end{align}
Let $m^*$ be the maximum $m$ over the course of the circuit and $k^*$ be the $k$-th gate where $m^*$ is achieved
\begin{align}
     m^* := \max\limits_{k \in[J]}m_{k}\hspace{2mm}, k^* := \mathop{\mathrm{argmax}}\limits_{k \in [J]} (m_k), 
\end{align}
If for simplicity we assume $\|O\|=1$ and that variations in the error made by the Riemann sum approximation as a function of $\delta$ and $k$ are sub-leading to the other effects we consider, we then have the following result for the maximum number of unique Pauli terms at any given point in the evolution (assuming that $\delta$ is chosen such that the evolution is in the regime where $\nmax$ has not reached the worst case scaling of $4^n$ for an $n$-qubit system):
\begin{equation} 
\label{eq: N_max bound}
\nmax \approx \frac{2-m^*}{m^*}\, \frac{\| O_{k^*} \|^2}{\delta^{m^*}(1 -\delta^{2-m^*})},
\end{equation}

\begin{figure}
    \centering
        \includegraphics[width=0.7\linewidth]{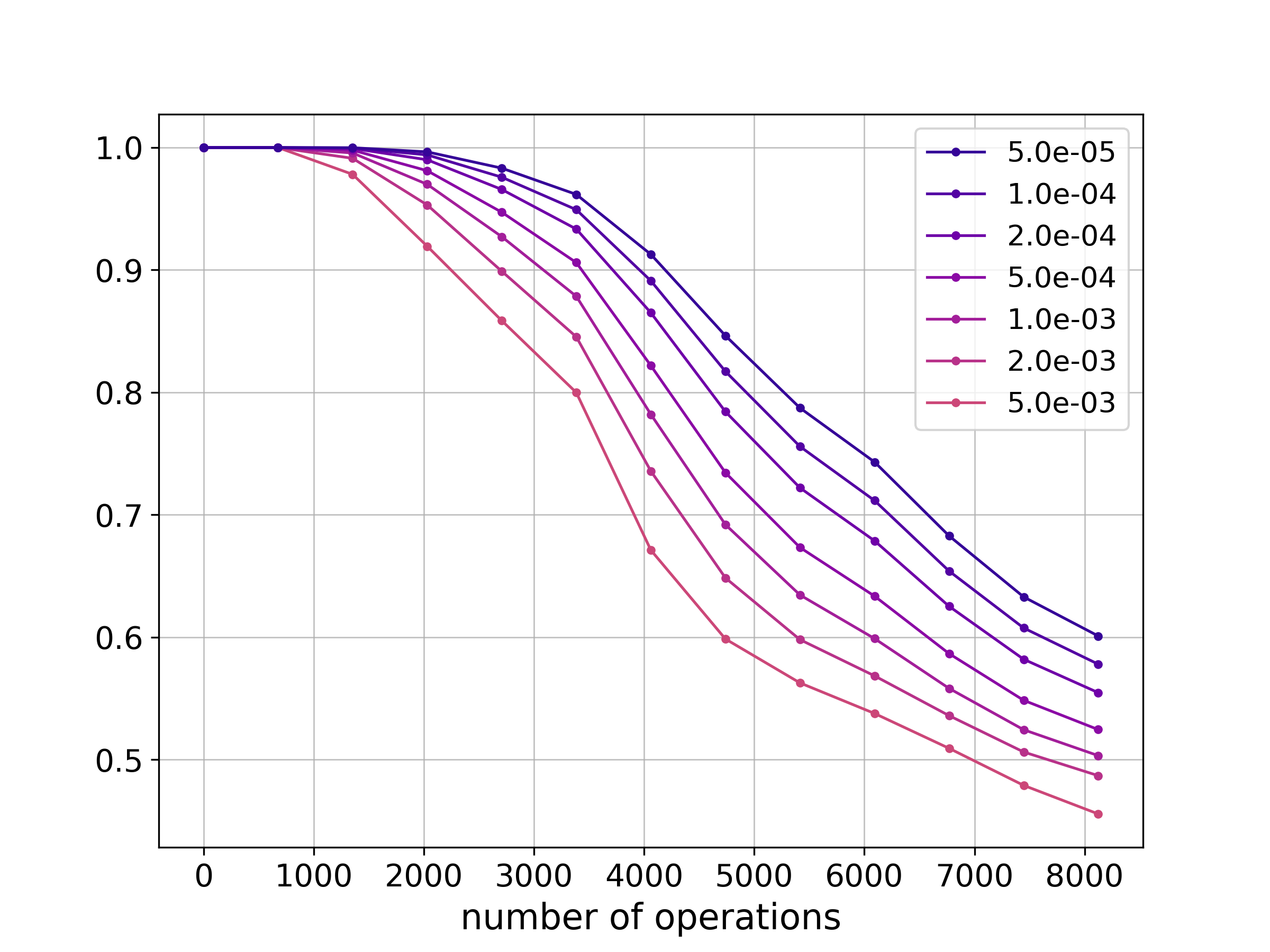}
        \label{fig: loss of mass rx_eq_pi_by_4.}
    \caption{Decrease in the norm $\|O_k\|$ of the evolved observable over the execution of the circuit for various values of truncation threshold. The circuit was constructed with $\theta_X$ sampled uniformly from $[\pi/4,\pi/4]$ and $T = 30$.}
    \label{fig:Ok_norms_random_rx}
\end{figure}

An important consequence of \cref{eq: N_max bound} is that gaps between Pauli-growth curves such as those shown in \cref{fig:pauli-growth-and-varying-m} can be shown to be approximately regular with respect to variations in $\delta$. In particular, we have that
\begin{align} 
\label{eq: N_max_gaps}
\log\left(\frac{N_{\max}(\delta_1)}{N_{\max}(\delta_2)}\right)
\approx m^* \log\left(\frac{\delta_2}{\delta_1}\right) 
+ 2\log\left(\frac{\|O_{k^*}(\delta_1)\|}{\|O_{k^*}(\delta_2)\|}\right)
+ \log\left(\frac{1 - \delta_2^{2-m^*}}{1 - \delta_1^{2-m^*}}\right).
\end{align}
When $\delta_1$ and $\delta_2$ are small, \cref{eq: N_max_gaps} can be further approximated as follows:
\begin{align}
\label{eq: N_max_gaps_small_delta}
\log\left(\frac{N_{\max}(\delta_1)}{N_{\max}(\delta_2)}\right)
\approx m^* \log\left(\frac{\delta_2}{\delta_1}\right)
+ 2\log\left(\frac{\|O_{k^*}(\delta_1)\|}{\|O_{k^*}(\delta_2)\|}\right)
\end{align}
As shown in \cref{fig:Ok_norms_random_rx}, when variations in $\delta$ are small so are the variations in the norm $\|O_k\|$ and hence we expect the behavior of the LHS of \cref{eq: N_max_gaps_small_delta} to be mostly characterized by the first term on the RHS. In other words, uniform (multiplicative) changes in $\delta$ approximately lead to uniform (multiplicative) changes in  $\nmax$.

We end this section by noting that since $\|O_k\| \leq \|O\| = 1$, \cref{eq: N_max bound} can be written as:
\begin{align}
    \label{eq:N_k_simplified}
    \nmax &\lesssim \frac{2-m^*}{m^*}\,  \frac{1}{\delta^{m^*}(1-\delta^{2-m^*})},
\end{align}
where we use $\lesssim$ to denote an upper bound up to a constant factor. We also note that using the bound shown in \cref{eq:N_k}, we can estimate other (absolute) moments of the coefficient distribution. For any $l \geq 1$, we get: 
\begin{equation}
\sum_{P \in \mathcal P_k} |c_P^{(k)}|^l \approx \frac{(2-m_k)}{(l-m_k)}\frac{(1-\delta^{(l-m_k)})}{(1-\delta^{(2-m_k)})} \|O_k\|^2.
\end{equation}\label{eq:higher_moments}
These estimates also show the dependence of the resources required to compute the evolved observable $O_k$ on the power-law exponent $m$. For example, the fourth-moment is called 
\textit{generalized Pauli purity} and is related to magic resources; see Theory Box 6 in \cite{Rudolph:2025gyq} and the references therein for more details.

\subsection{A resource efficient method to estimate $\nmax$}\label{subsec: N_max_in_practice}
In this section, we present numerical results demonstrating that \cref{eq: N_max_gaps,eq: N_max_gaps_small_delta} serve as a useful tools for predicting memory requirements at finer resolution. Directly using \cref{eq: N_max bound} for instance can prove to be challenging since this requires accurate estimates of the power law exponent $m$, which as we show in \cref{app:m_estimation_challenges} can prove to be a non-trivial problem. Furthermore, any relative error in $m$ leads to a relative error in $\nmax$ multiplied by $|\log(\delta)|$.\footnote{For a $5 \%$ error in estimating $m$ and a $\delta=5\times10^{-5}$ this could potentially mean a $50\%$ error in the $\nmax$ prediction. }

\begin{figure}
    \centering
    \includegraphics[width=0.7\linewidth]{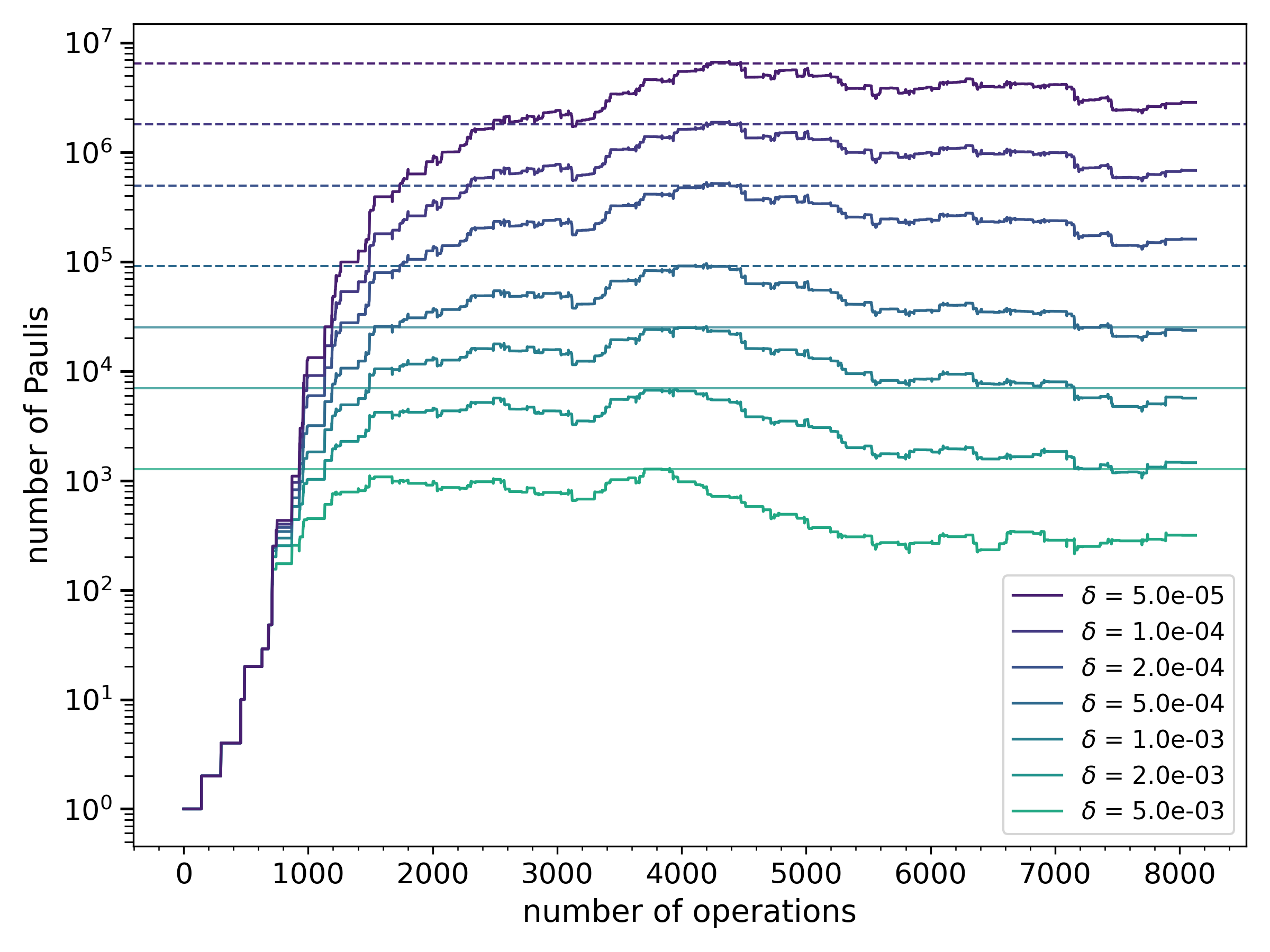}
    \caption{Estimating $\nmax$ via extrapolation : The exact values of $\nmax$ corresponding to the three largest values of $\delta$ (solid horizontal lines) are used to extrapolate and predict $\nmax$ for the smaller values of $\delta$ indicated in the plot. The predictions are within 6\% of the true $\nmax$ values. The circuit used is constructed with each $\theta_X$ sampled uniformly from $[-\pi/4,\pi/4]$ and $T = 30$.}
    \label{fig:N_max by interpolation}
\end{figure}

Importantly, while our theoretical analysis provides insight into this observed regularity, we show that the logarithmic consistency enables a powerful numerical bootstrapping approach: we can extrapolate simulation requirements from coarser truncation thresholds (larger $\delta$) to predict resource needs at finer thresholds (smaller $\delta$), dramatically reducing the computational cost of resource estimation. In practical terms, a scientific user would run this Pauli curve for very coarse thresholds at a cheap price and successively decrease $\delta$ by some constant ratio $r$. If the curves look similar then one can use this ratio $r$ to extrapolate $\nmax$ with a simple linear regression. Empirically, we find that getting just a few data points with $\delta_0 \approx 0.005, r = 1/\sqrt{2}$ is a good starting point. For \cref{fig:N_max by interpolation}, we obtained $\hat{N}_{\text{max}}(\delta)$ in less than a minute on a 6-core CPU with less than 6\% error on the $\nmax$ obtained by running the full Pauli Path simulation.

As another check of the predictive power of this method, we also apply this resource estimate to the 2D Ising dynamics on an $11\times11$ square lattice with open boundary conditions studied in Ref.~\cite{PRXQuantum.6.020302}. In particular, we implement the first order Trotterization (Eq. 19 of \cite{PRXQuantum.6.020302}) with $h = 3.044382$, $t = 0.92$, and $\Delta t = 0.04$ and perform PPS runs with 5 values of $\delta$ starting from $\delta_0 = 2^{-12}$ with $r = 1/\sqrt{2}$. Our generated estimate for $\nmax$ is $\sim 2.1$ billion and $48.5$ hours for the runtime\footnote{The extrapolation in runtime can be done using the simple power law scaling as a function of $\delta$ that we explore in \cref{sec:convergence}}. The reported value $\nmax$ in Fig. 8 of Ref.~\cite{PRXQuantum.6.020302} appears to be roughly $2.5$ billion. We note that the estimation procedure takes just under $7$ min on a single CPU core, providing valuable runtime and memory information. This yields a ballpark estimate even though this particular example results in the distribution of the absolute values of the coefficients exhibiting larger deviations from a power law form than that shown in for instance in \cref{fig:coeff-distribution}. For a discussion of these deviations, see~\cref{app:large_deviations}.

We end this section by pointing out some potential sources of errors that can skew such $N_{\mathrm{max}}$ estimates. The first is that the tail of the distribution contributes significantly to the second moment estimate in \cref{eq: second moment estimate} (as in the inset of \cref{fig: coeff distribution pi_by_6} where we provide an explicit numerical example of this effect). If the tail then also exhibits large deviations from a power law this might skew $N_{\mathrm{max}}$ estimates. Second, if $N_{\mathrm{max}}$ occurs close to the end of the circuit in the test runs, then the estimates can be less accurate in some cases as we only have a partial picture of the Pauli growth curves. (We speculate this is related to the small but visually apparent horizontal ``drift'' of $N_{\mathrm{max}}$ from one $\delta$ to the next as seen for example in \cref{fig: growth of paulis}).

\section{Scrutinizing convergence of PPS}
\label{sec:convergence}
Notably, PPS suffers from a lack of useful error bounds as $\delta$ decreases. The question of convergence as a function of the truncation parameter $\delta$ has been mentioned in other works (most notably in \cite{Begusic:2023jwi}), but mostly with a focus on finding the largest (cheapest) $\delta$ approximation which achieves a certain desired result.

We flip the investigation around and seek to understand when this classical method appears reliable and where it points to its own shortcomings. Even when PPS fails to converge, it can still be useful as an independent estimate to hardware runs, especially in the age of quantum devices that require sophisticated error mitigation techniques that need tuning to extract answers with confidence.

As in the last section, the main protocol we propose is tracking the convergence of a given expectation value for finer truncation parameter, starting with a rather coarse $\delta$. This gives a way to start with modest computational resources before choosing to commit to a large simulation which may or may yield a reliable answer. We find that the simulation time scales as a power law in $1/\delta$, leading to an efficient way to estimate the CPU time needed to extend to smaller $\delta$.

For notational convenience, we define a few variables. $\delta_0$ an initial, relatively large truncation parameter, $r$ is the ratio between successive $\delta_n$, $t_{\text{CPU}}$ is the maximum wall time. Let $\mathcal{O}_n$ denote the observable expectation value with truncation parameter $\delta_n$ and $t_n$ denote the simulation time for computing $\mathcal{O}_n$. Let $\varepsilon_{\text{tol}}$ denote a convergence tolerance.

The protocol can then be summarized succinctly as follows. Compute successive estimates $\mathcal{O}_n$ using $\delta_n = r^n\delta_0 $ and proceed until $t_n$ exceeds $t_{\text{CPU}}$ or $\ell$ successive estimates of $\mathcal{O}_n$ converge to within $\varepsilon_{\text{tol}}$. Because of the power law scaling of time vs $\delta$, one can also predict how long experiments with larger $\delta_n$ will take.

Since PPS is a frontier classical simulation method, it necessarily makes some problems look easy while struggling with other. One goal of this paper is to bifurcate quantum simulation problems by looking at convergence patterns. In the first category are problems where taking successively small $\delta_n$ leads to apparent convergence. Here, we use the term apparent convergence since there are no guarantees that finer and finer resolutions will yield the same value, unless all terms are kept. There may be many small-scale fluctuations so that the technical definition of convergence may very sensitively depend on $\varepsilon_{\text{tol}}$. However, in those cases it may still lead to high confidence in the resulting answer.

The second category is composed of problems for which, under finite $t_{\text{CPU}}$, the approximate expectation value fluctuates wildly, and estimates can no longer be trusted to have a high level of accuracy. Surprisingly, some of the results reproducing expectation values the IBM experiment~\cite{Kim:2023bwr} belong to this class. Without the hardware result, PPS applied to those problems would not yield a reliable answer.
However, if the PPS answer was calculated before the quantum result, it would still lead to a guiding range of where the true expectation should lie. 

In both cases, the simulation time scales as a power law in $1/\delta$ (for sufficiently small $\delta$). Thus, given finite computational resources, one can always check convergence based on coarser $\delta$ and use those runtimes and $\nmax$ values to extrapolate the resource requirements for pushing to finer $\delta$.

As a preview of this section, we highlight some surprising results on the utility of PPS.
\begin{enumerate}
    \item  For some instances where PPS reproduced published IBM results, it looks far from converged.
    \item  Even when unconverged, PPS may still be useful in tandem with results from quantum hardware, for example, serving as a Monte Carlo-like estimate. 
    \item  Exact simulation is obtained as $\delta \rightarrow 0$, but simulations with smaller $\delta$ do not always get closer to the exact answer. 
    \item  Conversely, extending a given circuit to larger depth may actually lead to PPS converging with \textit{fewer resources}. 
\end{enumerate}

\subsection{Apparent Convergence of PPS}
\label{subsec:apparent_convergence}
We now turn our attention to studying the convergence of expectation values with PPS. As a classical approximation method, there is inherently some level of precision sacrificed. However, it is important to note that even with a quantum device, the presence of errors, whether from gate fidelities, measurement readout or simply statistical shot noise, there are always error bars present in the final quantum result.

\begin{figure}[h!tb]
    \centering
   \includegraphics[width=\textwidth]{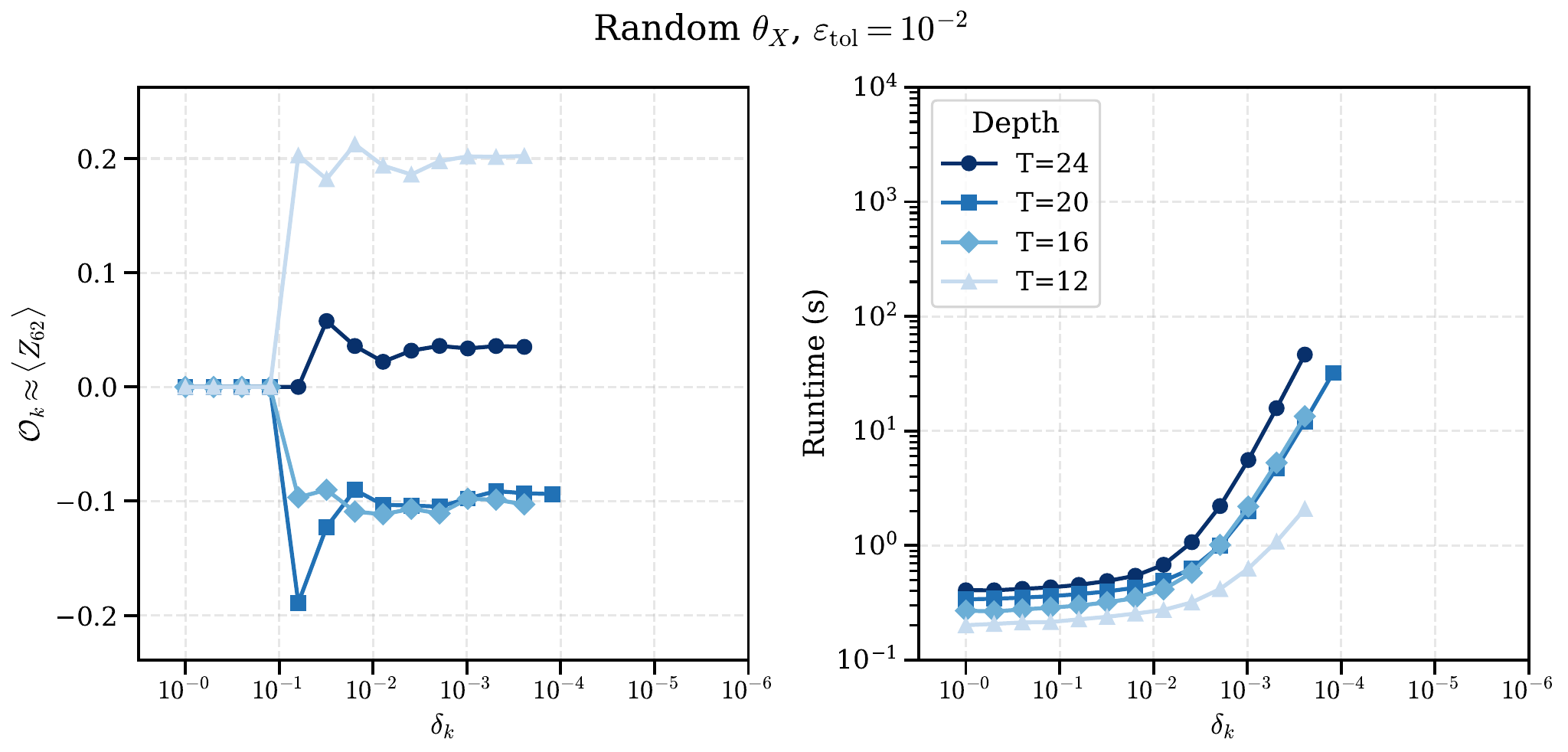}
    \includegraphics[width=\textwidth]{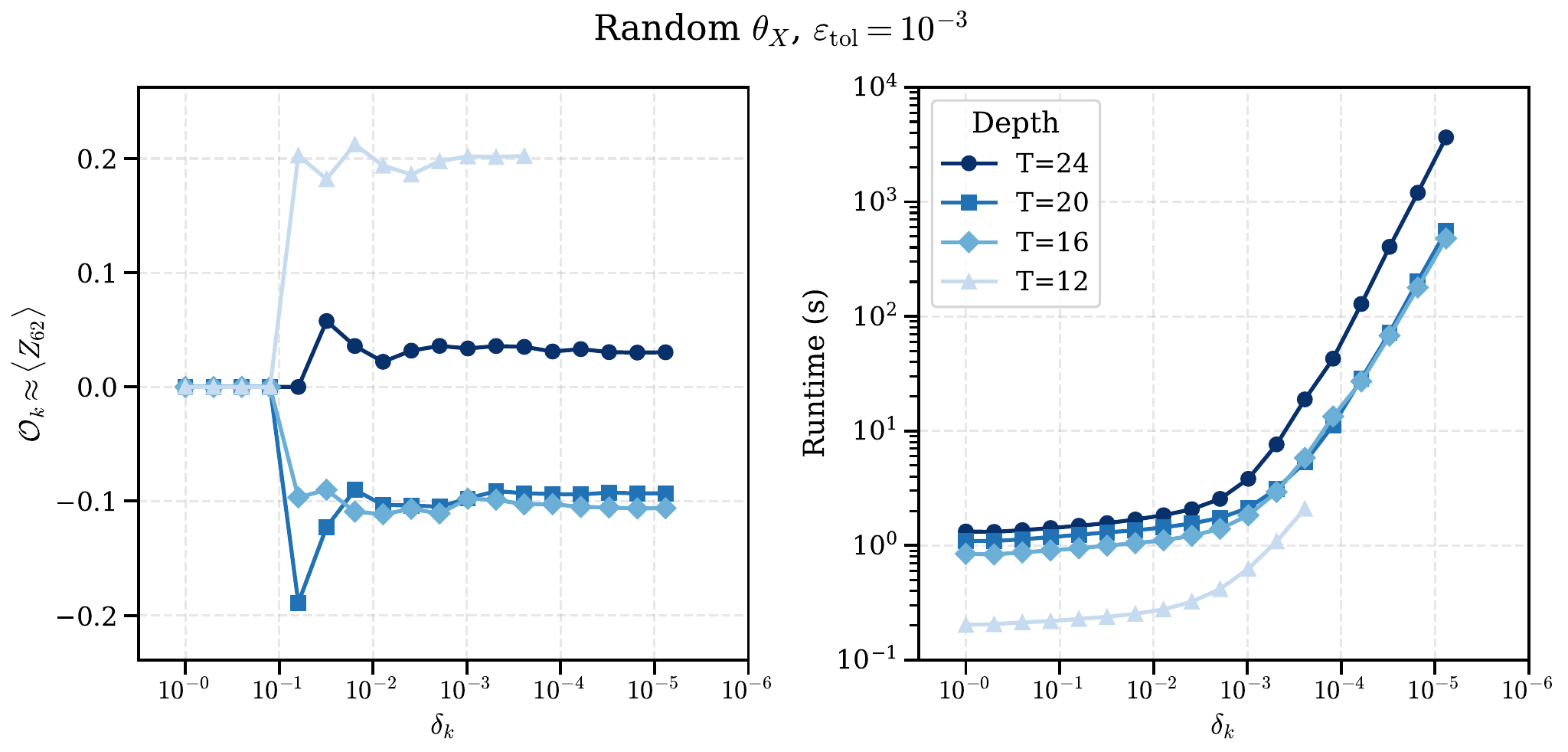}
    \caption{Convergence plots showing $\mathcal{O}_k$, the approximation to the expectation value $\langle Z_{62} \rangle$, as a function of $\delta_k$ with randomly sampled $\theta_{X}$. The top row is with $\varepsilon_{1}= 0.01$ and the bottom row is for a stricter $\varepsilon_{2}= 0.001$. We vary the number of Trotter steps $T$ within the range $[12,24]$. The right column has the runtimes demonstrating a power law scaling with a transition in the exponent. Notably, $T=16$ and $T=20$ have nearly identical runtimes. With $T=12$, the successive points converged at the same value of $\delta_{k}$ for both $\varepsilon_{1}$ and $\varepsilon_{2}$. The offset in runtimes on the bottom row (for $T=16, 20, 24$) are due to using a 64 CPU core vs a single CPU core due to smaller $\delta_{k}$. }
    \label{fig:z_62_T_2}
\end{figure}

It may then be tempting to simply ask for convergence of PPS within error bars comparable to those present from quantum devices. Because convergence depends very sensitively on $\varepsilon_{\text{tol}}$ and the ratio between successive $\delta_n$, $r$, there is some trial and error needed. Additionally, this is further complicated since the error induced by the truncation is non-monotonic.

\begin{figure}[h!tb]
    \centering
    \includegraphics[width=\textwidth]{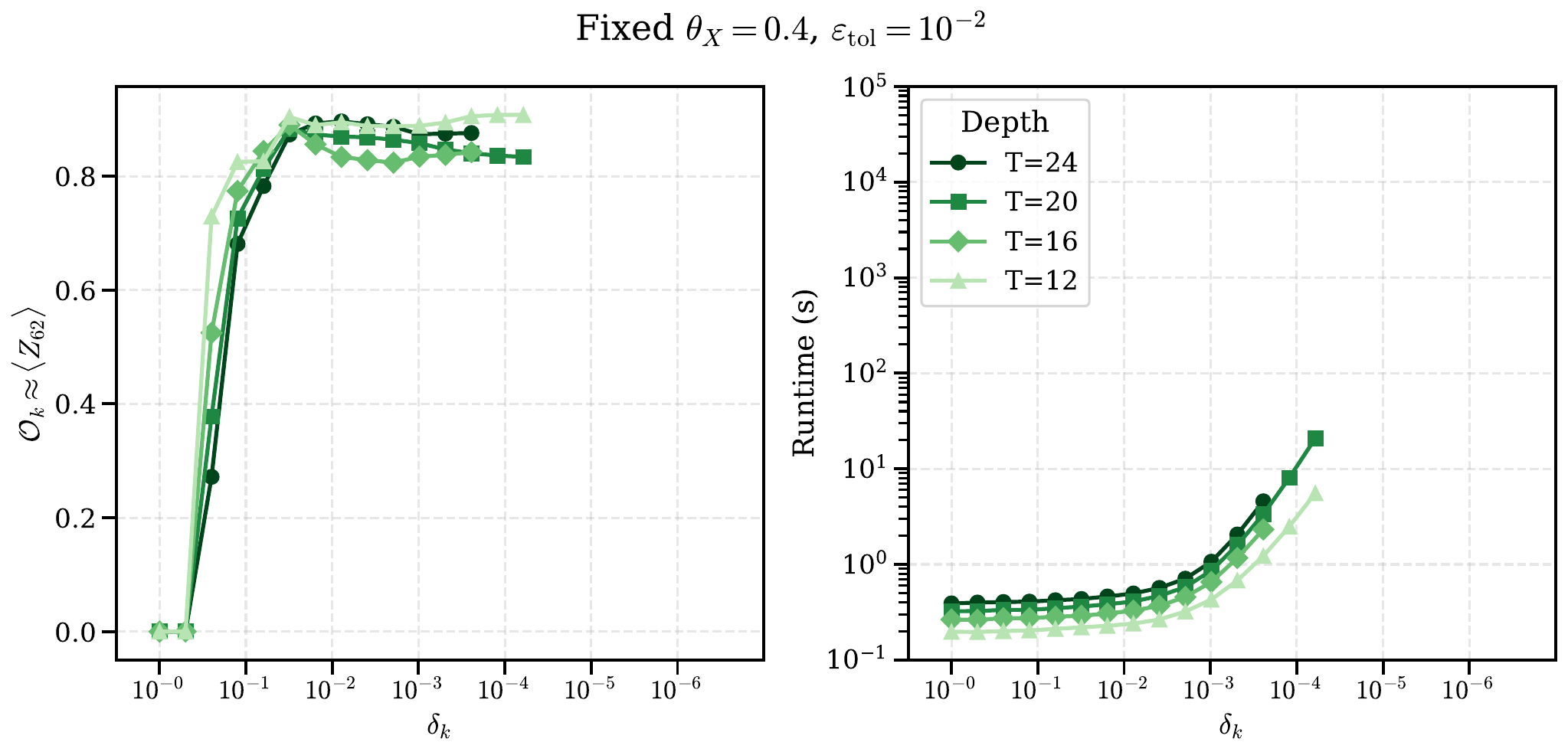}
    \includegraphics[width=\textwidth]{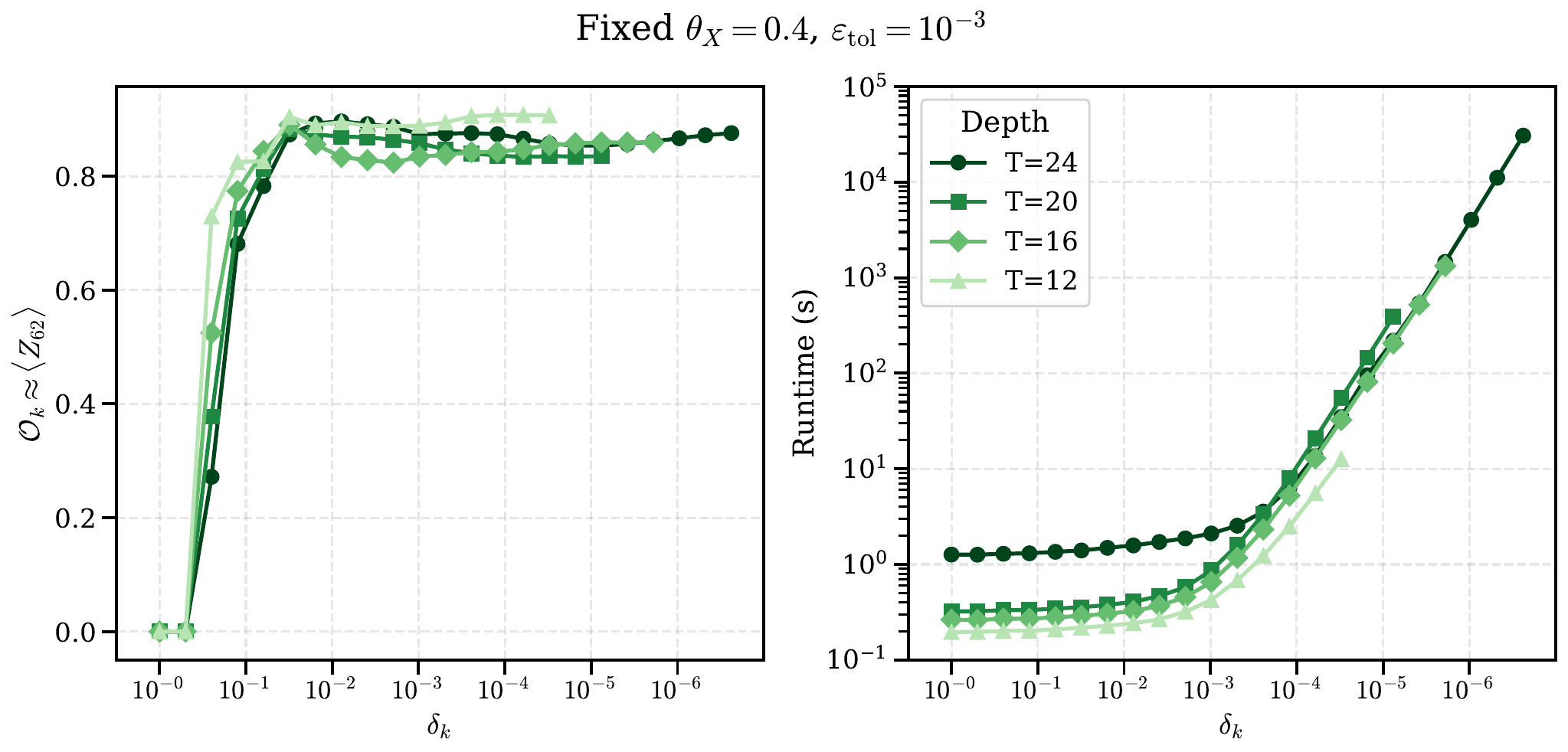}
    \caption{Convergence plots of $\mathcal{O}_k$ with fixed $\theta_{X} = 0.4$. We vary the number of Trotter steps $T$ within the range $[12,24]$. For $\varepsilon_{\text{tol}}=10^{-2}$ (top row), all experiments converge within $20$s on a single CPU core. Changing to a stricter $\varepsilon_{{\text{tol}}}=10^{-3}$, the estimated expectation value does not change significantly, but takes significantly longer to converge. The offset of $T=24$ in the bottom row is again due to using a 64 core CPU.}
    \label{fig:z_62_T_correlated}
\end{figure}

As an example, we apply our protocol to the kicked quantum dynamics with the single site observable $\langle Z_{62} \rangle$. For random single qubit rotations and varying number of Trotter steps, we see that extending PPS beyond $\delta=10^{-3}$ does not significantly change the estimate. See~\cref{fig:z_62_T_2}.

There are a number of interesting observations about this convergence behavior. First, we see the sensitivity to $\varepsilon_{\text{tol}}$. For the same choice of $\delta_0, r, t_{\text{CPU}}$, decreasing $\varepsilon_{\text{tol}}$ from $10^{-2}$ to $10^{-3}$, one needs to decrease $\delta$ by almost 2 orders of magnitude. Correspondingly, the largest $t_n$ at $T=24$ was over 100 times larger without officially converging, even when using 64 CPU cores vs a single core\footnote{For large $\delta$, a single core is faster due to communication overhead.}. Based on this clear power law scaling of the time, we can extrapolate that the next two points would take 20 hours and 50 hours, respectively.

Here, the extra computational resources simply confirm the confidence in the convergence, but do not change the result\footnote{Strictly speaking, the experiment for $T=24$ did not converge when taking $\ell=3$, pointing to the trial and error nature of defining convergence criteria.}.
Decreasing $\delta$ in principle should approach the correct answer since $\delta \rightarrow 0 $ corresponds to exact simulation. In practice, the lack of control on the signs of the corrections highlights a key feature about using PPS. Picking a smaller $\delta$ does not always yield a more useful result. If running for one day gives the same level of approximation as running it for 1 minute, then cheaper is better.

Another salient aspect of the convergence in~\cref{fig:z_62_T_2} is that a deeper circuit does not necessarily imply that one needs a smaller $\delta$ to achieve comparable accuracy than the $\delta$ needed for the shorter circuit. Compare the case of $T=12$ versus $T=24$ for $\varepsilon_{\text{tol}}= 10^{{-2}}$. Although the circuit size has effectively doubled, the same $\delta_{n} \approx 10^{-3.5}$ is sufficient to get a converged estimate. Thus, every circuit must be probed independently! One cannot easily extrapolate convergence difficulty across different system sizes or trotter steps or even different rotation angles of the same circuit architecture.

Next, we turn to the more widely studied form of the dynamics \cref{eq:kicked-ising} where the single qubit rotations are all correlated and equal to each other.

Here the details of the observable convergence are slightly different from the prior random angle case. However, many of the findings carry over. For example, we see that the largest fluctuations are already captured by the looser criteria of $\varepsilon_{\text{tol}}=10^{-2}$. For the shallowest circuits, we converge very quickly regardless of the system size, but for the deepest circuits, there are small fluctuations which cause the convergence procedure to fail for $\varepsilon_{\text{tol}}=10^{-3}$.
We continue to observe the power law extrapolation of $t$ versus $\delta$. There is a significant transition in the runtime scaling around $\delta \sim 10^{-3}$.

As discussed previously, there is a strong dependence of convergence on $\varepsilon$, and we can see that the observable as a function of $\delta$ is non-monotonic. Thus, depending on the spacing $r$, you may encounter local minima, which falsely indicate convergence. For this set of circuits and this observble, around $\delta \approx 10^{-3}$  however, there appear to be fewer inflection points with large fluctuations. Additionally, due to the estimates being non-monotonic, there are no guarantees that convergence prevents larger fluctuations as $\delta$ continues to decrease.

\subsection{Beyond Apparent Convergence}

PPS has been used to successfully reproduce expectation values consistent with quantum hardware experiments at modest $\delta$~\cite{Begusic:2023jwi}. However, applying our proposed convergence procedure yields an inconclusive result about the predicted value without priors from quantum hardware. The lack of convergence in these cases highlights the limitations on the utility of PPS as both a discovery and a verification tool.

PPS works best when there are fewer Pauli terms to track or when the truncated terms are significantly smaller than the surviving ones. Clifford circuits provide an extreme limit, mapping Pauli terms to single Paulis, so PPS is exact. The presence of magic \cite{Dowling:2024wvo,Rudolph:2025gyq}, or non-Cliffordness, thus is one proxy for the difficulty of PPS methods\footnote{The name Sparse Pauli Dynamics (SPD) comes from this intuition}. Indeed, when $\theta = \pi/4$, the branching in Eq.~\ref{eq:branching-and-merging} yields equal weights.

In this section, we again study circuits from the quantum kicked Ising model. This time, we fix $T=20$ and instead vary the single qubit rotation angles $\theta_X$ between $0.3$ and $0.7$, slowly dialing up the magic or non-Cliffordness.

\begin{figure}[h!tb]
    \centering
    \includegraphics[width=\textwidth]{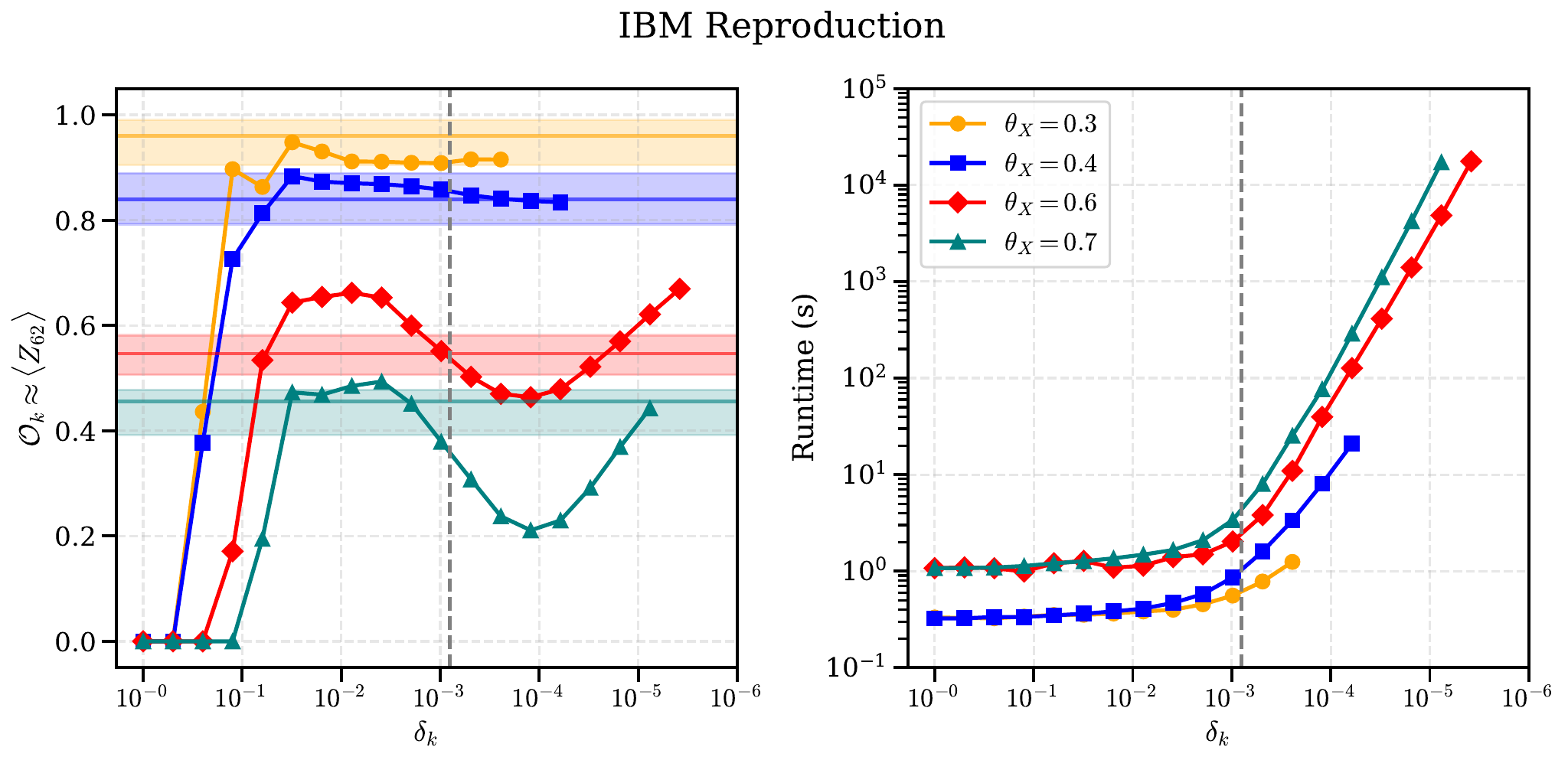}
    \caption{Convergence plots of $\mathcal{O}_k$ for $\theta_{X} = 0.3, 0.4, 0.6, 0.7$. We fix the number of Trotter steps $T=20$ within the range. For $\varepsilon_{\text{tol}}=10^{-2}$ (top row), the experiments corresponding to $\theta_{X} = 0.3, 0.4$ both converge within $20$s on a single CPU core. However, the experiments corresponding to $\theta_{X} = 0.6, 0.7$ exhibit much larger fluctuations with decreasing $\delta_k$ and did not converge within the $10$ hour limit we set on compute time for any single $\delta_k$. We note that the latter pair of experiments were carried out on a 64 core CPU. The vertical dashed line indicates the truncation parameter used by Garnet as the single $\delta$ that approximates the quantum result across all angles.}
    \label{fig:z_62_rx}
\end{figure}

For $\theta_X = 0.3, 0.4$, we find that the observable converges with $\varepsilon_{\text{tol}} = 0.01$ within seconds as in the previous sections. However, as $\theta_X$ increases to $0.6,0.7$, we see that there are larger fluctuations even past $\delta = 10^{-3}$. This is the marked contrast from the previous convergence of observables, well into the regime that the power law scaling of runtime kicks in. The wide variance in values reported reflects the lack of reasonable error bounds for smaller delta. We also plot the publicly available data from the experiments in Ref.~\cite{Kim:2023bwr}.  See \cref{fig:z_62_rx}.

The point we emphasize is that PPS here does not readily yield a prediction. One can estimate a range of values where the true expectation value is expected to live, but no single value stands out as more ``accurate'' a priori. Additionally, the lack of controlled error means that the final answer could live outside the range of previously computed estimates.

With a criteria of getting $\ell = 3$ successive points to agree within $\varepsilon_{\text{tol}}$, it's clear at least two more data points are required. Extrapolating based on the power law scaling, we determine that the two additional data points for $\theta_{X} = 0.6$ would require over $11$ and $44$ hours respectively and the two additional data points for $\theta_{X} = 0.7$ would require over $17$ and $70$ hours respectively. If the goal is to obtain a self-consistent answer rather than finding a $\delta$ which matches a known result, this would dramatically shift the narrative around the necessary classical resources. Rather than minutes on a laptop, it may require several days on a cluster.

The lack of convergence here also highlights the fact that a smaller $\delta$ does not translate in principal to a more accurate answer. Thus, it may be more beneficial to the user to perform a more coarse approximation with the understanding that the error bar might be somewhat significant.

\section{Discussion}
Pauli Path Simulation represents one of a handful of frontier classical methods for quantum simulation, offering unique capabilities for tackling utility-scale quantum circuits. Our work provides quantum researchers with a systematic framework for understanding when PPS delivers reliable estimates and where its fundamental limitations emerge.

\textbf{Resource Prediction and Mathematical Structure} One of our key contributions reveals the surprisingly elegant mathematical structure underlying PPS dynamics. A power-law distribution \cref{eq: power_law} can be used to approximately model the distribution of the absolute values of the Pauli coefficients corresponding to the evolved observable. This observation enables ballpark predictions of computational resource requirements using only brief and computationally inexpensive test runs. This discovery transforms resource estimation from guesswork into a principled extrapolation procedure, allowing researchers to predict memory and runtime scaling for finer truncation parameters $\delta$ without committing to expensive full simulations. 

\textbf{Counterintuitive Performance Characteristics} Our convergence analysis reveals several surprising aspects of PPS behavior that challenge conventional intuition. Reducing the truncation parameter $\delta$ does not always improve accuracy, and deeper quantum circuits may actually converge more easily than shallower ones—effects we trace to the degree of non-Cliffordness in the evolution. These findings underscore the critical need for problem-specific benchmarking rather than attempting to extrapolate convergence difficulty across different system sizes, circuit depths, or parameter regimes. Each quantum simulation problem must be probed independently using our proposed convergence protocol.

\textbf{Testing for Apparent Convergence} Our systematic study reveals two distinct operational regimes for PPS. In the first regime, problems exhibit apparent convergence where successive refinements in $\delta$ lead to stable expectation value estimates with modest computational resources. In the second regime, PPS reports large fluctuations that do not readily yield a single answer. Surprisingly, some problems that successfully reproduce published IBM results actually fall into this second category when examined through our convergence lens, highlighting the distinction between post-hoc verification and independent prediction capability.

\textbf{Practical Utility Beyond Apparent Convergence} Even when PPS fails to achieve formal convergence, it retains significant practical value as a complementary tool to quantum hardware experiments. In the era of noisy intermediate-scale quantum devices, where error mitigation requires careful tuning of numerous parameters—noise models, gate selections for Pauli twirling, and mitigation protocols -- PPS provides an independent classical estimate that can guide optimization efforts. For instance, if quantum results exhibit opposite signs from the entire PPS convergence trajectory, this signals potential issues with error mitigation parameters that warrant investigation.

The method proves particularly valuable in applications where approximate answers suffice. In the Sample-based Quantum Diagonalization method introduced in \cite{Robledo-Moreno:2024pzz}, the precise energy estimate may matter less than capturing the correct amplitude weights for subspace population. Similarly, quantum machine learning applications often use expectation values as inputs to classical networks, where classification accuracy depends more on distinguishing between different examples than on precise numerical values. In these contexts, PPS serves as an inexpensive classical pre-training tool, with subsequent refinement possible on quantum hardware.

PPS also presents an economically attractive first solution.
Since it is primarily CPU-based, experiments are significantly less expensive than GPU or QPU simulations. This can accelerate the research pipeline, providing insight into whether a problem is classically difficult and how much additional computational resources is needed to push to finer resolutions.

\textbf{Implications for Quantum Advantage} As quantum computing advances toward demonstrating practical quantum advantage, distinguishing between problems that succumb to improved classical methods and those requiring genuine quantum resources becomes increasingly critical. Our systematic framework for evaluating PPS—one of the most promising frontier classical simulation techniques—helps sharpen these boundaries. By providing reliable diagnostics for when classical simulation suffices versus when quantum hardware offers genuine advantages, this work contributes to the broader effort of mapping the quantum-classical computational landscape.

Rather than positioning this analysis as criticism of PPS, we view it as essential guidance for a powerful but nuanced simulation tool. The method's relative novelty necessitates careful characterization of its capabilities and limitations. Our convergence protocol offers an inexpensive preliminary assessment that helps researchers determine whether their specific quantum simulation problem lies within the realm of classical tractability or genuinely requires quantum resources. This diagnostic capability becomes increasingly valuable as we approach the threshold of quantum computational advantage, providing a principled approach to distinguishing between classical and quantum regimes in contemporary quantum simulation challenges.

\bibliographystyle{alpha}
\bibliography{refs}

\pagebreak
\appendix
\section{PPS Convergence with BlueQubit Backend}\label{app:code}

This section contains the code snippet to use the BlueQubit backend for studying PPS convergence. After using qiskit to build the circuit of interest, one simply needs to specify the observable as well as successive $\delta_{k}$. With relatively few runs, one can also perform the regression to extrapolate the time needed to go to finer $\delta$. The code snippet shown below can be used to reproduce the PPS curve corresponding to $\theta_X = 0.3$ in \cref{fig:z_62_rx}

\begin{codeblock}
import bluequbit
import matplotlib.pyplot as plt
import numpy as np

# helper function that returns all the nearest neighbors on 
# IBM's 127 qubit heavy hex lattice
from bluequbit.library.helpers.hardware_connectivites import IBM_127_HEAVY_HEX_MAP

# helper function that returns qiskit Pauli objects given the
# qubits idxs for the X,Y, and Z operators in the Pauli string
from bluequbit.library.helpers.pauli_sum import construct_pauli_from_idx_lists
from qiskit import QuantumCircuit

# you can get the BlueQubit API key after signing up at https://app.bluequbit.io
bq = bluequbit.init("<BLUEQUBIT API KEY>")

num_qubits = 127
num_trotter_steps = 20

rzz_angle = -np.pi / 2
rx_angle = 0.3

# list of the coefficient thresholds (delta) in decreasing order
deltas = [1 / 2**i for i in range(13)]

# list to store the expectation value <Z_62> for each choice of delta
expectation_values = []

# construct circuit for this pair of rzz and rx angles
qc = QuantumCircuit(num_qubits)
for _ in range(num_trotter_steps):
    for edge in IBM_127_HEAVY_HEX_MAP:
        qc.rzz(rzz_angle, edge[0], edge[1])
    for i in range(num_qubits):
        qc.rx(rx_angle, i)

for delta in deltas:
    # set the Pauli path coefficient threshold
    options = {
        "pauli_path_truncation_threshold": delta,
    }

    # run PPS
    expectation_values.append(
        bq.run(
            qc, device="pauli-path", pauli_sum=pauli_sum, options=options
        ).expectation_value
    )

plt.plot(-np.log10(deltas), expectation_values)
\end{codeblock}

\section{Bit efficient representation of Pauli sums}\label{app:implementation}

We note that in the Pauli decomposition of the observable $O_k$, each Pauli $P\in\mathcal{P}_k$ admits the following symplectic representation:
\begin{align}
    P = (-i)^{\alpha_P} X^{x_{P}^{(1)}} Z^{z_{P}^{(1)}} \otimes X^{x_{P}^{(2)}} Z^{z_{P}^{(2)}} \otimes \cdots \otimes X^{x_{P}^{(n)}} Z^{z_{P}^{(n)}},
\end{align}
where $\alpha_{P}\in\mathbb{Z}$, $X$ and $Z$ denote the usual Pauli $X$ and $Z$ matrices respectively, and for each qubit $\ell$ we have the boolean pair $(x_{P}^{(\ell)}, z_{P}^{(\ell)}) \in \mathbb{Z}_2^2$. We define the boolean vector $\nu_{P} := \left( z_{P}^{(1)},\dots,z_{P}^{(n)},x_{P}^{(1)},\dots,x_{P}^{(n)} \right) \in \mathbb{Z}_2^{2n}$. Thus, to efficiently represent the observable $O_k$ in memory we can store two length $|P_{k}|$ arrays that contain the set of complex coefficients $\{c_P^{(k)}\}_{P\in\mathcal{P}_k}$ and integers that generate the phase factors $\{\alpha_P\}_{P\in\mathcal{P}_k}$, as well as a $|P_{k}|\times(2n)$ array such that each row contains a vector of the form $\nu_{P}$\footnote{In practice, we use the idea presented in \cite{Begusic:2023thb} to encode each vector $\nu_{P}$ into an array of 64 bit unsigned integers and hence the $2D$ array consisting of all such vectors can be stored as a $|P_{k}|\times(2\lfloor n/64 \rfloor)$ array}.

As shown for instance in \cite{PRXQuantum.6.020302}, all the operations required to perform this evolution in the Heisenberg picture can be efficiently performed by manipulating the three arrays. For example, we can check the commutativity of two Pauli operators as follows. Let $\nu_{P}$ and $\nu_{\sigma_{k+1}}$ be the boolean vectors corresponding to the Pauli operators $P,\sigma_{k+1}$ respectively. We define the quantity $s_{P\sigma_{k+1}}$ as follows:
\begin{align}
    s_{P\sigma_{k+1}} := \mathrm{parity}(\nu_P[n:] \hspace{1mm} \& \hspace{1mm} \nu_{\sigma_{k+1}}[:n]) \oplus \mathrm{parity}(\nu_P[:n] \hspace{1mm} \& \hspace{1mm} \nu_{\sigma_{k+1}}[n:]),
\end{align}
where $(\cdot) \hspace{1mm} \& \hspace{1mm} (\cdot)$ represents the bitwise $\mathrm{AND}$ operator,  $(\cdot) \oplus (\cdot)$ represents the $\mathrm{XOR}$ operation (addition modulo 2), $\nu_P[n:]$ and $\nu_P[:n]$ represent the first and last (respectively) $n$ components of the vector $\nu_P$, and $\mathrm{parity}(\cdot)$ returns the number of non-zero bits modulo 2. Upon computing $s_{P\sigma_{k+1}}$, the commutativity of $P$ and $\sigma_{k+1}$ can be determined as follows:
\begin{align}
    s_{P\sigma_{k+1}} = 0 &\implies [P,\sigma_{k+1}] = 0 \\
    s_{P\sigma_{k+1}} = 1 &\implies \{P,\sigma_{k+1}\} = 0.
\end{align}
Hence, at every stage in the evolution, we can construct the sets $\mathcal{P}^{comm}_{k}, \mathcal{P}^{anti}_{k}$ by computing $s_{P\sigma_{k+1}} \forall P \in \mathcal{P}_k$. Products of the form $P' = \sigma_{k+1}P$ can also be efficiently constructed by using the symplectic representation in the following way:
\begin{align}
    \alpha_{P'} &= \alpha_P + \alpha_{\sigma_{k+1}} + 2 \cdot \mathrm{count}\left(\nu_P[:n] \hspace{1mm} \& \hspace{1mm} \nu_{\sigma_{k+1}}[n:]\right) \\
    \nu_{P'} &= \nu_{P} \oplus \nu_{\sigma_{k+1}},
\end{align}
where $\mathrm{count} \hspace{1mm} (\cdot)$ returns the number of non-zero bits. Upon evolving the observable by all the unitaries in the set $\{U_j\}_{j\in[J]}$, the required expectation value is then given by
\begin{align}
    \langle O \rangle = \bra{0} O_J \ket{0} = \sum_{P\in\mathcal{P}_J}c_P^{(J)}\bra{0}P\ket{0} = \sum_{\substack{P\in\mathcal{P}_{J} \\ P = Z-\mathrm{type}}}c_P^{(J)}(-i)^{\alpha_P},
\end{align}
where the final sum only runs over the $Z$-type Paulis in $P_J$, i.e. Paulis for which the boolean vectors are of the form $\nu_P = \left( z_{P}^{(1)},\dots,z_{P}^{(n)},0,\dots,0 \right)$.

\section{Coefficient distribution for correlated angles}
\label{app:correlated_distributions}

\begin{figure} 
    \centering
    \begin{subfigure}{1.0\linewidth}
        \centering
        \includegraphics[width=1\linewidth]{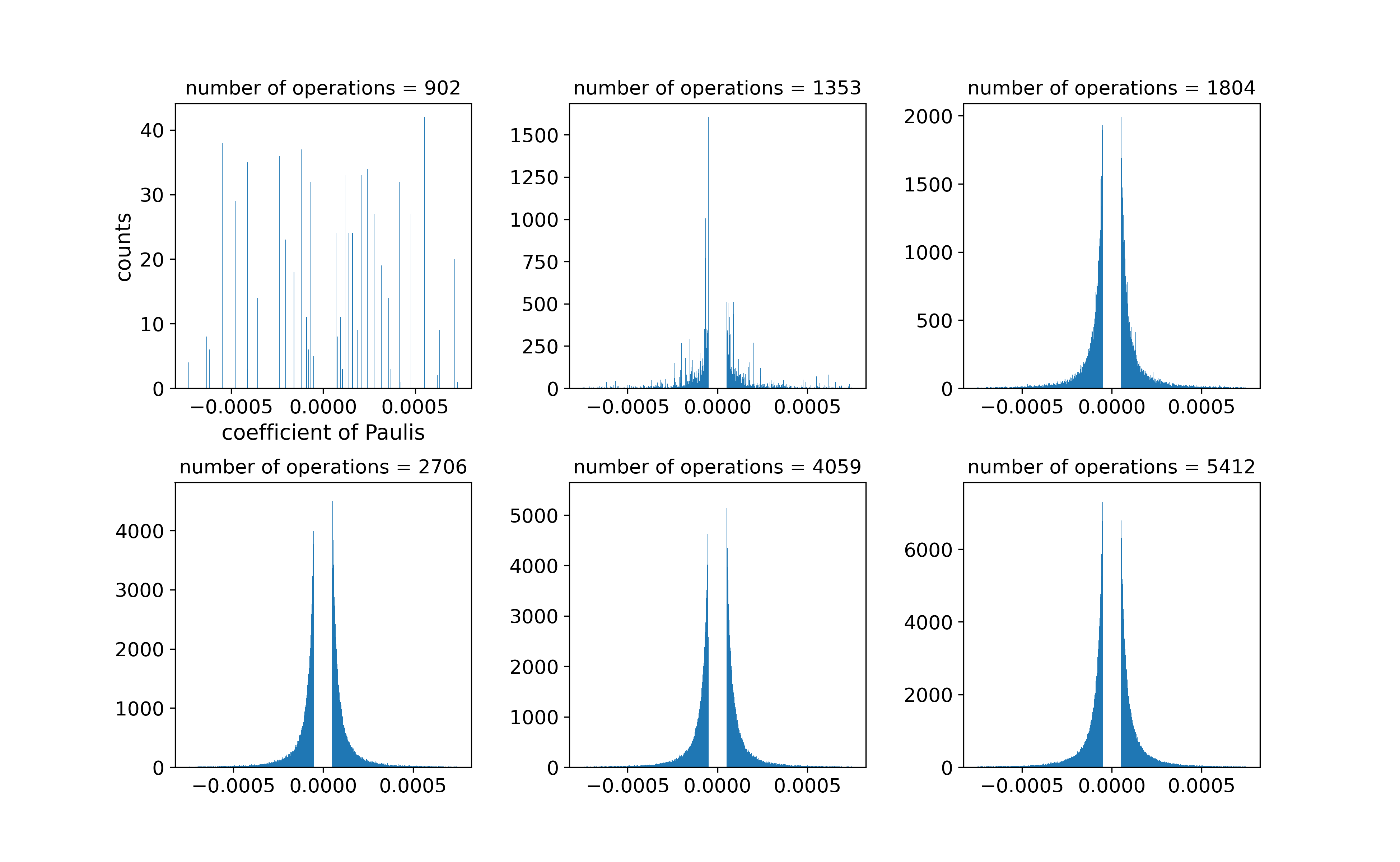}
        \caption{Evolution for circuit with $\theta_X$ equal to $\pi/6$ and $T = 20$ for $\delta = 5\times10^{-5}$.}
        \label{fig: invariance of distribution}
    \end{subfigure}
    \centering
    \begin{subfigure}{1.0\linewidth}
        \centering
        \includegraphics[width=1\linewidth]{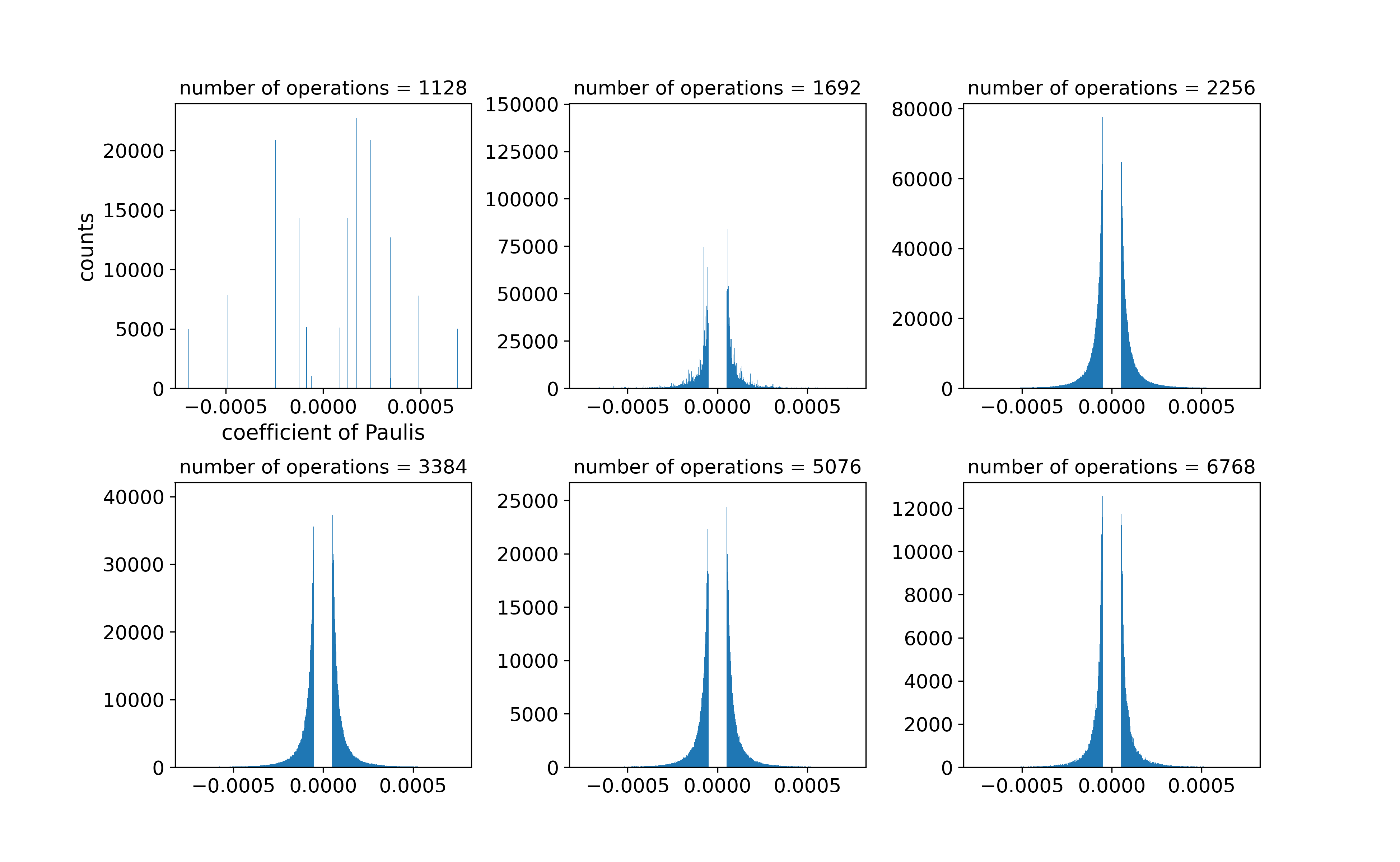}
        \caption{Evolution for circuit with $\theta_X$ equal to $\pi/4$ and $T = 25$ for $\delta = 5\times10^{-5}$.} 
        \label{fig:evolution of distribution (pi_by_4)}
    \end{subfigure}
    \caption{Evolution of the distribution of the Pauli coefficients of the evolved observable for circuits with correlated angles}\label{fig:evolution_of_distribution}
    
\end{figure}

\cref{fig:coeff-distribution} shows that the distribution of the absolute values of the Pauli coefficients converges to a power law over the course of circuit evolution under the kicked Ising dynamics with $\theta_X$ chosen at random. \cref{fig:evolution_of_distribution} show that this feature is not restricted to completely random angles and can also be observed for fixed, correlated $\theta_X$ as well.

\section{Analytical results on Pauli Proliferation }\label{app:analytical}

For any Pauli operator $P$, with associated data $(c_P, \alpha_P, \nu_P)$, the scalar $(-i)^{\alpha_P - \mathrm{sum}(\nu_P[ :n] \,\& \,\nu[n: ])}\times c_P$ is always real, and we shall abuse notation by denoting this quantity again by $c_P$ (and refer to them as `real coefficients' if necessary). These are the actual coefficients of the underlying Pauli string operators in the linear expansion of the observable in the Pauli basis. Consider the observable $O_k$ defined in \cref{eq:pauli_decomp_k}, generated by applying the first $k$ unitary gates of the circuit shown in \cref{eq:circuit_unitary}, followed by coefficient truncation at each step.
 We assume that the real coefficients $c_{P}^{(k)}$ are distributed according to a PDF $\rho_k: \mathbb{R} \rightarrow \mathbb{R}_{\geq 0}$:
\begin{align}
    \label{eq: prob_dist}
    \mathrm{Pr}(\alpha \leq c_{P}^{(k)} \leq \beta) = \int_\alpha^\beta \rho_k(x)dx.
\end{align}
By imposing a $\delta \in \mathbb{R}_{\geq 0}$ truncation threshold, we introduce the following constraint on $\rho_k$:
\begin{align}
    \rho_k(x) = 0 \hspace{2mm} \forall |x| < \delta.
\end{align}

We now ask the following question: How can we use $\rho_k$ to determine the new PDF $\rho_{k+1}$ used to describe the distribution of the numbers $c_{P}^{(k+1)}$ appearing in the Pauli decomposition of $O_{k+1}$ obtained by applying the gate $U_{k+1} = e^{-i(\theta/2)\sigma}$ to $O_k$ followed by truncation? For the purpose of answering this question, we may assume without loss of generality that the gate $U_{k+1}$ is non-Clifford. We can compute an approximation to $\rho_{k+1}$ by making the following simplifying assumptions:

\begin{hypo}(PPS hypothesis)
\label{hypothesis}
\begin{enumerate}
\item[i.)] The distribution satisfied by the numbers $c_{P}^{(k)}$ corresponding to Pauli operators in $\mathcal{P}_k^{comm}$ and those corresponding to Pauli operators in $\mathcal{P}_k^{anti}$ are the same, i.e. \cref{eq: prob_dist} holds with the same PDF $\rho_k$ $\forall P\in\mathcal{P}_k^{comm}$ and $P\in\mathcal{P}_k^{anti}$.
\item[ii.)] 
For the Pauli gate generator $\sigma$, define the set of Pauli operators $\sigma\mathcal{P}^{anti}_{k} \subseteq \mathcal{P}_{k+1}$ as follows:
\begin{align}
    \sigma \mathcal{P}^{anti}_{k} := \{\sigma P \hspace{1mm} | \hspace{1mm} P \in P^{anti}_{k}\}.
\end{align}
The distribution of coefficients of Paulis in $\mathcal P_k^{anti}$ induces by restriction a distribution on coefficients of Paulis belonging to $\mathcal P_k^{anti} \cap \sigma \mathcal P_k^{anti}$. Then, for any randomly chosen $P \in \mathcal P_k^{anti} \cap \sigma \mathcal P_k^{anti}$, we assume the following independence for the pair $\{P, \sigma P\} \subset \mathcal P_k^{anti} \cap \sigma \mathcal P_k^{anti}$:
\begin{align}
    \mathrm{Pr}\left(\alpha \leq c_P^{(k)} \leq \beta,\, \alpha' \leq c_{\sigma P}^{(k)} \leq \beta'\right) = \mathrm{Pr}\left(\alpha \leq c_P^{(k)} \leq \beta\right) \times \mathrm{Pr} \left(\alpha' \leq c_{\sigma P}^{(k)} \leq \beta'\right)
\end{align}
\end{enumerate}
\end{hypo}

\begin{remark}
\label{remark}
An important consequence of these assumptions is the following: Let  $\mathcal P_{k+1} := \mathcal P_k^{comm} \cup (\mathcal P_k^{anti} \cup \sigma_{k+1}\mathcal{P}^{anti}_{k})$ be the new set of Pauli terms prior to truncation, and suppose $P' \in \mathcal P_{k+1}$ belongs to $\mathcal P_k^{anti} \cap \sigma \mathcal P_k^{anti}$. Let $P' = i\sigma P$ be such that \cref{eq:branching-and-merging} reads $c^{(k+1)}_P = c_P^{(k)}\cos(\theta) + c_{P'}^{(k)} \sin(\theta)$. Then  $$\mathrm{Pr}(\alpha \leq c^{(k+1)}_P \leq \beta) = \mathrm{Pr}(\alpha \leq  c_P^{(k)}\cos(\theta) + c_{P'}^{(k)} \sin(\theta) \leq \beta) = \int_\alpha^\beta (\rho_k)^\star_\theta (t) dt,$$ where \begin{equation}(\rho_k)^\star_\theta (t) :=  |\sec(\theta)| \int_{-\infty}^\infty \rho_k(u)\rho_k(t\sec(\theta)-u\tan(\theta)) du \label{eq:rho_conv_definition}\end{equation} is the convolution. 
\end{remark}

Let us now estimate the number of Paulis remaining in $\mathcal P_{k+1}$ after $\delta$-truncation is applied to $\mathcal P_{k+1}$. Let $\varphi = |\mathcal P_k^{anti}|/|\mathcal P_k|$ be the fraction of Paulis in $\mathcal P_k$ that anticommute with $\sigma$. And let $\eta = |\mathcal P_k^{anti} \cap \sigma \mathcal P_k^{anti}|/|\mathcal P_k|$ be the fraction of the anticommuting Paulis in $\mathcal P_{k+1}$ that already exists in $\mathcal P_k$. From \cref{eq:branching-and-merging} the number of Paulis in $\mathcal P_k^{anti} \setminus \sigma\mathcal P_k^{anti}$, after truncation, gets reduced by a factor of $2\int_{\delta\sec \theta} ^\infty \rho_k(t) dt.$ Likewise, $\sigma\mathcal P_k^{anti} \setminus \mathcal P_k^{anti} $ gets reduced by a factor of $2\int_{\delta\csc \theta} ^\infty \rho_k(t) dt$ and $\mathcal P_k^{anti} \cap \sigma \mathcal P_k^{anti}$ gets reduced by the factor 2$\int_\delta^\infty (\rho_k)^\star_\theta (t) dt$.

Therefore, after the gate application and truncation, the number of Paulis $N_k := |\mathcal P_k|$ changes as \begin{equation} N_k \mapsto N_k \bigl(1-\varphi +  2(\varphi-\eta)\bigl[\int_{\delta\sec \theta} ^\infty \rho_k(t) dt + \int_{\delta\csc \theta} ^\infty \rho_k(t)  dt \bigr] + 2\eta \int_\delta^\infty (\rho_k)^\star_\theta (t) dt  \bigr)    =: N_k'. \label{num_paulis_change} \end{equation}

Note that $N_k'$ is an approximation to $N_{k+1}$. Let us write $p(\theta) =  2\int_{\delta\sec \theta} ^\infty \rho_k(t) dt$, $q(\theta) = 2\int_{\delta\csc \theta} ^\infty \rho_k(t) dt$, and $r(\theta) =  2\int_\delta^\infty (\rho_k)^\star_\theta (t) dt$. Given the PPS hypothesis and the associated remark, for any $P \in \mathcal P_{k+1}$ we have

\begin{align}  
\mathrm{Pr}  (\alpha \leq c^{(k+1)}_P \leq \beta)   = \;& \bigl[(\mathrm{Pr}(P \in \mathcal P_k^{comm})  + \mathrm{Pr}(P \in \mathcal P_k^{anti}\setminus \sigma\mathcal P_k^{anti}) + \mathrm{Pr}(P \in \mathcal  \sigma\mathcal P_k^{anti} \setminus  \mathcal P_k^{anti}) \nonumber \\& + \mathrm{Pr} (P \in \mathcal P_k^{anti} \cap \sigma \mathcal P_k^{anti}) \bigr] \times \mathrm{Pr} (\alpha \leq c^{(k+1)}_P \leq \beta). \\  =\; & \frac{(1-\varphi) N_k}{N_k'} \mathrm{Pr} (\alpha \leq c_P^{(k)} \leq \beta) + \frac{(\varphi - \eta) N_k p(\theta) }{N_k'} \mathrm{Pr} (\alpha \leq c_P^{(k)}\cos(\theta) \leq \beta) \nonumber \\ &+ \frac{(\varphi - \eta) N_k q(\theta)}{N_k'} \mathrm{Pr} (\alpha \leq c_P^{(k)}\sin(\theta) \leq \beta) + \frac{\eta N_k}{N'_k}\mathrm{Pr}(\alpha \leq  c_P^{(k)}\cos(\theta) + c_{P'}^{(k)} \sin(\theta) \leq \beta). 
\end{align}

Thus, we see that the probability distribution function $\rho_{k+1}$ of coefficient of Paulis in $\mathcal P_{k+1}$ (after truncation) is related to that of $\mathcal P_k$ by \begin{equation}\rho_{k+1}(t) = \frac{\bigl((1-\varphi) \rho_k(t) + (\varphi-\eta)[p(\theta)\sec(\theta)\,\rho_k(sec(\theta)t) + q(\theta)\csc(\theta)\,\rho_k(csc(\theta)t)] + \eta \overline{(\rho_k)}_\theta(t)\bigr)}{{(1-\varphi + (\varphi-\eta)(p(\theta) + q(\theta)) + \eta r(\theta))}}, \label{eq:density_evol}\end{equation}
where $\overline{(\rho_k)}_\theta(t)$ is the $\delta$-truncation of the convolution $(\rho_k)^\star_\theta(t)$ defined as $\overline{(\rho_k)_\theta}(t) = (\rho_k)_\theta^\star(t)$ if $|t| \geq \delta$, and $\overline{(\rho_k)}_\theta(t) = 0$ otherwise.

\begin{figure}[!htbp]
    \centering
    \includegraphics[width=0.5\linewidth]{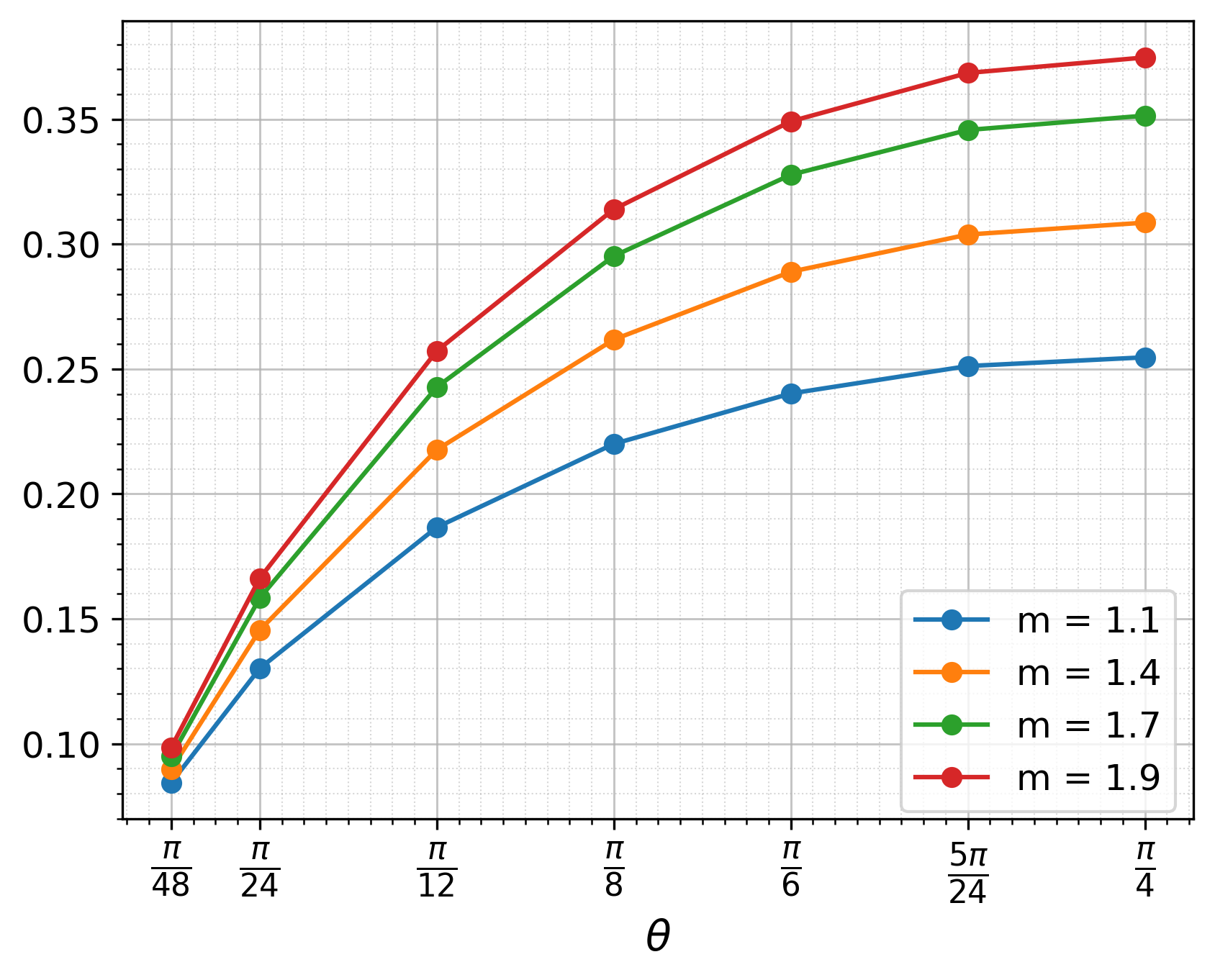}
    \caption{Plots showing the numerical value of $s(\theta) := \int_{-\delta}^{\delta} \rho_\theta^\star(t) dt$ for the convolution \cref{eq:rho_conv_definition} for different values of $m$. Under the PPS hypotheses, the set $\mathcal P^{anti} \cap \sigma \mathcal P^{anti}$ gets shrunk by the factor $1-s(\theta)$ after applying a $\delta$-truncation. This factor is denoted by $r(\theta)$ in the text, and it intervenes in \cref{eq:num_paulis_change_model}. Due to the symmetry satisfied by $\rho_\theta^\star$, $s(\theta)$ satisfies $s(\pi/4-\theta) = s(\pi/4+\theta)$.}
    \label{fig: rho conv integrals}
\end{figure}

Having derived the general equation governing the transformation of $\rho_k$, we now switch to the specific case of the power-law distribution \cref{eq: power_law}. For the power-law distribution, we have $p(\theta) = (\cos \theta)^m$ and $q(\theta) = (\sin\theta)^m$. While $r(\theta)$ does not seem to have a simple form, it can be computed effectively with numerical integration as shown in \cref{fig: rho conv integrals}. \cref{num_paulis_change} can now written as:
\begin{equation} N'_k = N_k (1-\varphi + (\varphi-\eta)((\cos\theta)^m + (\sin \theta)^m) + \eta r(\theta)). \label{eq:num_paulis_change_model} \end{equation}

\begin{figure}
    \centering
    \begin{subfigure}{0.475\textwidth}
        \centering
        \includegraphics[width=\linewidth]{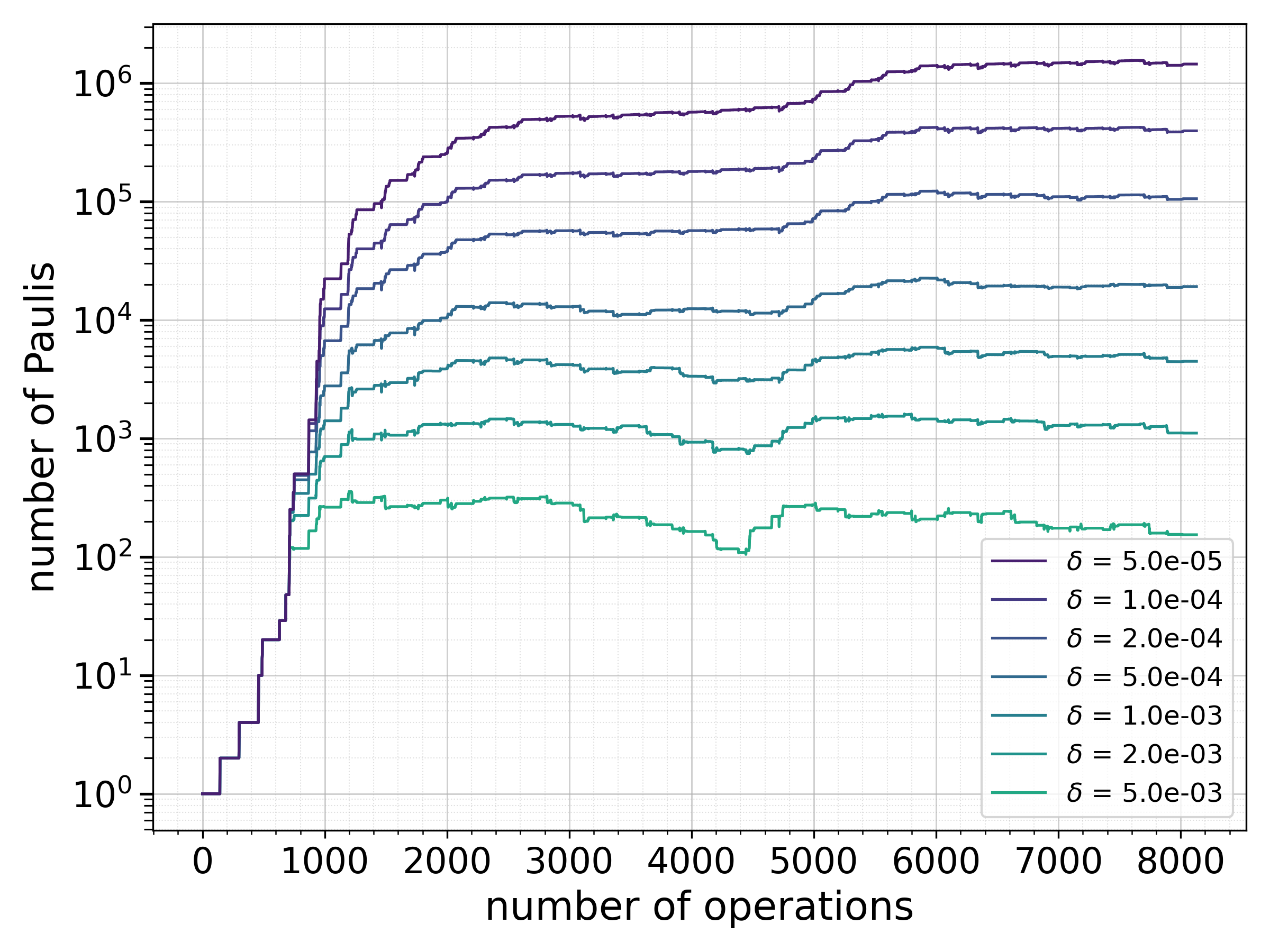}
        \label{fig:growth of Paulis rx pi_by_6}
    \end{subfigure}
    \begin{subfigure}{0.475\textwidth}
        \centering
        \includegraphics[width=\linewidth]{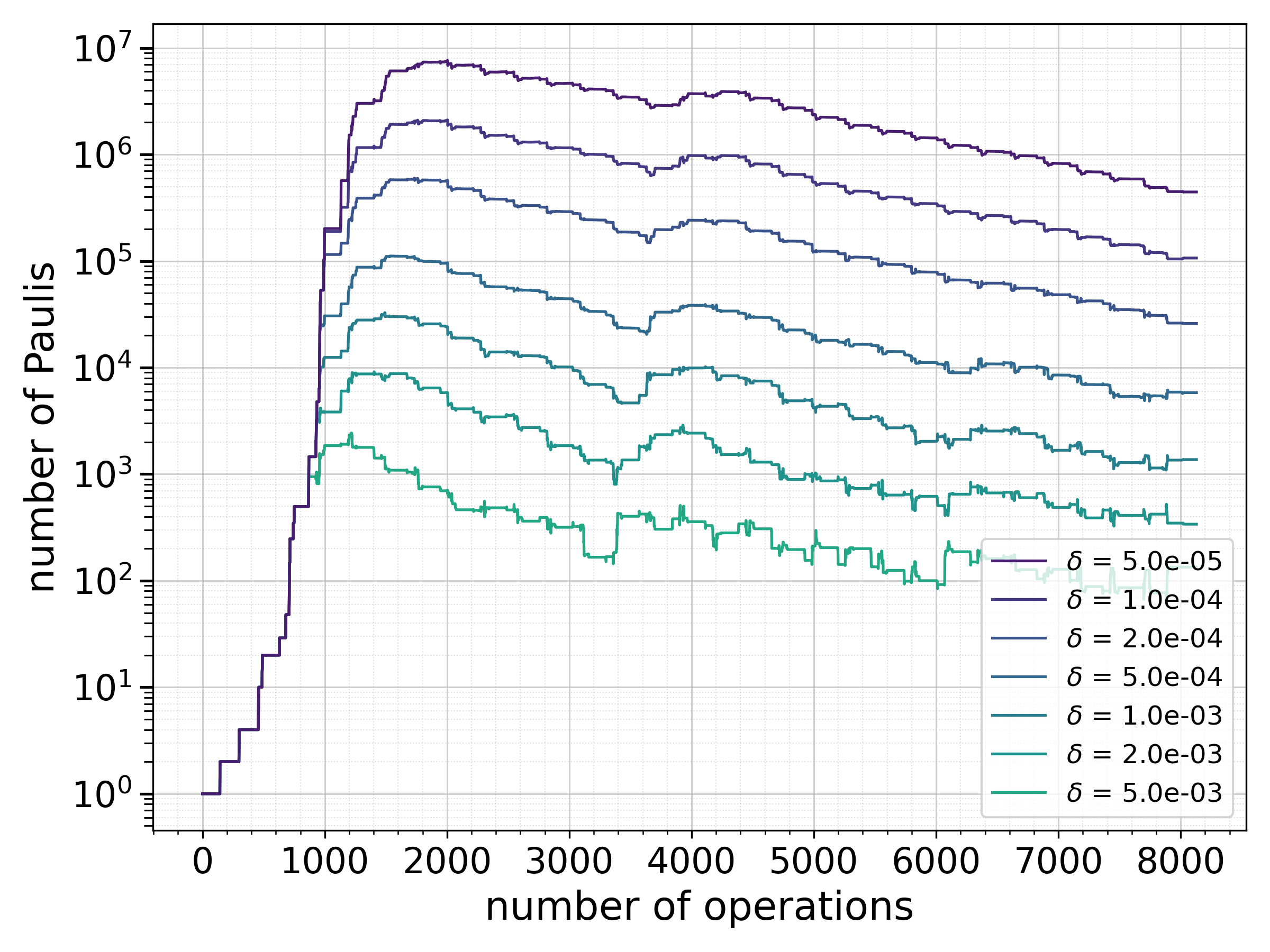}
        \label{fig: growth of paulis rx pi_by_4}
    \end{subfigure}
    \caption{Growth of Paulis over the execution of the circuit. Left: for circuit with $\theta_X$ equal to $\pi/6$. Right: for $\theta_X$ equal to $\pi/4$. We set $T = 30$ for both. A standout feature of these plots is the regularity of the growth curves with respect to $\delta$; the curves are almost identical for smaller values of $\delta$. This aspect of the simulation is explained by our modeling (see \cref{eq:num_paulis_change_model} and the subsequent discussion).}
    \label{fig: growth of paulis}
\end{figure}

\begin{figure}
    \centering
    \includegraphics[width=0.96\linewidth]{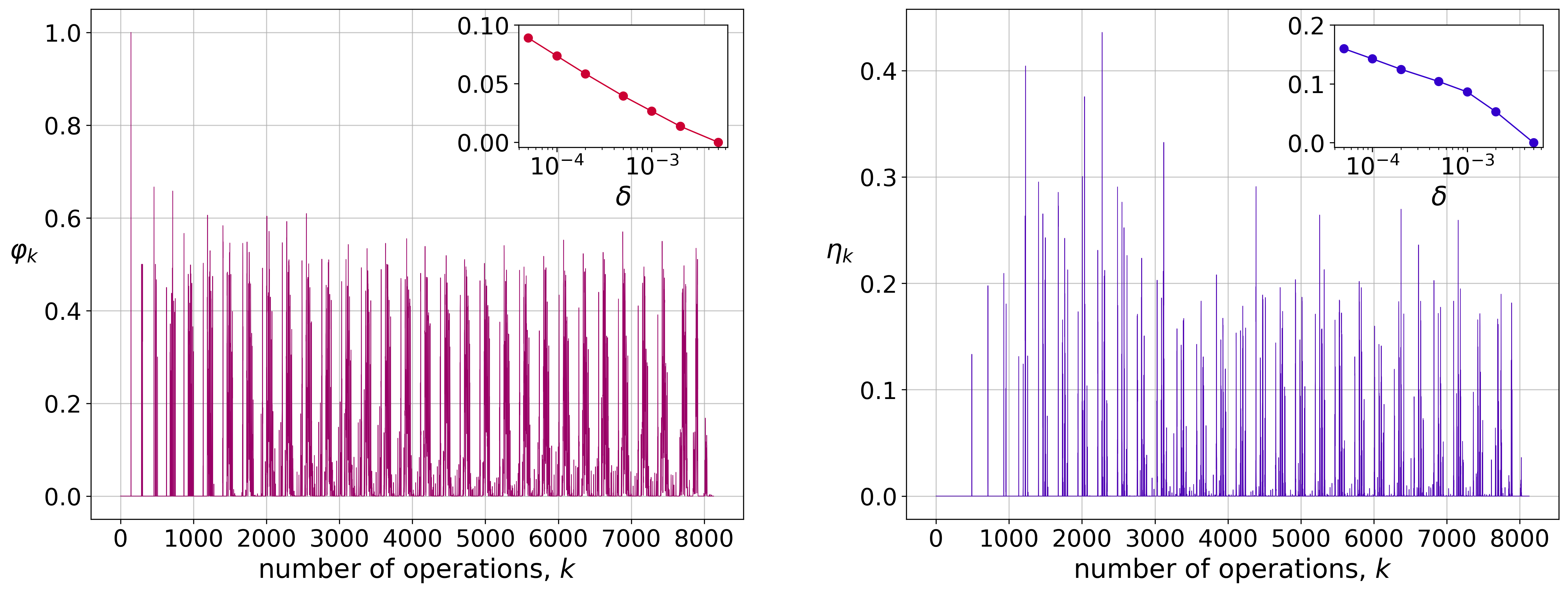}
    \caption{Fraction of anticommuting Paulis, $\varphi_k = |\mathcal P_k^{anti}|/|\mathcal P_k|$, and ``existing" anticommuting Paulis, $\eta_k =  |\mathcal P_k^{anti} \cap \sigma \mathcal P_k^{anti}|/|\mathcal P_k|$, plotted over the course of the circuit execution for the circuit constructed with each $\theta_X$ sampled uniformly from $-[\pi/4,\pi/4]$, $T = 30$ steps. The simulation is performed using $\delta = 5\times10^{-5}$. We see that $\eta$ spikes a few times in each Trotter step. There is an $\eta$-spike at $k = 2036$, and we show its effect  on the distribution in \cref{fig:eta wiggle random rx}. Inset, left:  Cosine distance (defined as  1 $-$ cosine similarity) between $\vec\varphi(\delta)$ and $\vec\varphi(\delta_0)$ as $\delta$ varies from $\delta_0 = 5\times 10^{-3}$ to $5\times 10^{-5}$. Here, we have interpreted the sequence $(\varphi_k)_{k=1, 2, \ldots, J}$ as a vector $\vec\varphi$. Inset, right: Variation of Cosine distance between $\vec\eta(\delta)$ and $\vec\eta(\delta_0)$. We observed in our experiments that the variation of $\vec\varphi$ and $\vec\eta$ with respect to $\delta$ is small and gradual.}
    \label{fig:ac fracs rx_eq_random}
\end{figure}

To maintain concise notation in this appendix, the dependence of $\varphi, \eta,$ and $m$ on $k$ is implied and not explicitly noted. We point out that the above recurrence model, which we derived using properties of the power-law distribution, captures the similarity of the Pauli-growth curves seen in \cref{fig: growth of paulis,fig:N_max by interpolation} (for large enough values of $k$, when the distribution is well-formed). The dependence of the model on $\delta$ arises implicitly through $m$, $\varphi$, and $\eta$. If there was no such dependence, \cref{eq:num_paulis_change_model} implies that the curves must be identical (translates). Numerical evidence, as shown in \cref{fig:pauli-growth-and-varying-m} and the insets of \cref{fig:ac fracs rx_eq_random}, suggests that the variation of $m$, $\varphi$, and $\eta$ with respect to $\delta$, though very subtle, is small and gradual. While these observations neatly explain why adjacent curves are so similar, they also suggest that \cref{eq:num_paulis_change_model} may be impractical for estimating the full Pauli growth curve.  

Let us now turn to the distribution itself. For the power-law model, \cref{eq:density_evol} which gives the changed density function reads \begin{equation} \rho_{k+1}(t) = \frac{([1-\varphi + (\varphi-\eta)((\cos\theta)^{2m} + (\sin\theta)^{2m})] \rho_k(t) + \eta\overline{(\rho_k)}_\theta(t))}{(1-\varphi +(\varphi-\eta)((\cos\theta)^m + (\sin\theta)^m)) + \eta r(\theta))}. \label{eq: density_change_full}
\end{equation}

Suppose for now that $\eta \ll 1$. Then, after ignoring the term $\eta \overline{(\rho_k)}_\theta(t)$ above, the changed density function is of the form
\begin{align}
    \rho_{k+1}(t) \approx \lambda  \rho_k(t)
\end{align}
for some constant $\lambda$ depending on $\varphi, \eta, \theta$ and $m$. Thus, under the PPS hypothesis and the assumption that $\eta \ll 1$, we see that the functional form of the density function (power law) remains invariant under the dynamics. In fact, imposing the normalization condition on $\rho_{k+1}$ implies that $\rho_{k+1} \approx \rho_k$. This is in fact consistent with numerical evidence.

The effect of the convolution term $\eta \overline{(\rho_k)}_\theta(t)$ on the dynamics is more subtle. Although we can ignore this term when $\eta \ll 1$ or when $\theta \approx 0$ (in which case $\overline{(\rho_k)}_\theta(t) \approx \rho_k(t)$) when considering a single step of the evolution, the cumulative effects over many gate applications can be significant. Furthermore, there could be many instances/gates when $\eta$ spikes to high values (resulting in large merges in the step $\mathcal P_k \rightarrow \mathcal P_{k+1}$).

\section{Deviations from a Power Law caused by $\eta$-spikes}\label{app:power_law_deviations}

In this section, we discuss two possible effects related to the $\eta$ dependent term in \cref{eq:num_paulis_change_model} that we have observed that lead to large visually apparent deviations of the distribution of the absolute values of the Pauli coefficients of the evolved observable from a power law. In particular these effects cause either transient or persistent deviations.

\subsection{Transient deviations}\label{app:eta-spikes}

For ease of notation, we shall denote the truncated convolution $\overline{(\rho_k)}_\theta$ simply by $\overline\rho_\theta$. A priori, the effect of the term $\eta\overline\rho_\theta$ in \cref{eq: density_change_full} is not obvious; it is also unclear how this term propagates and skews the coefficient distribution evolution throughout circuit execution. We do know, however, from numerical evidence (for the IBM circuits with Clifford RZZ gates that we have considered here) that the truncated power-law distribution is persistent and any perturbation to it seems transient. This begs the question: how to explain away the effect of $\eta \overline{\rho}_\theta(t)$ in \cref{eq: density_change_full} and retain the stability of the truncated power-law distribution for the dynamics in circuits considered thus far?

\begin{figure}
    \centering
    \includegraphics[width=0.96 \linewidth]{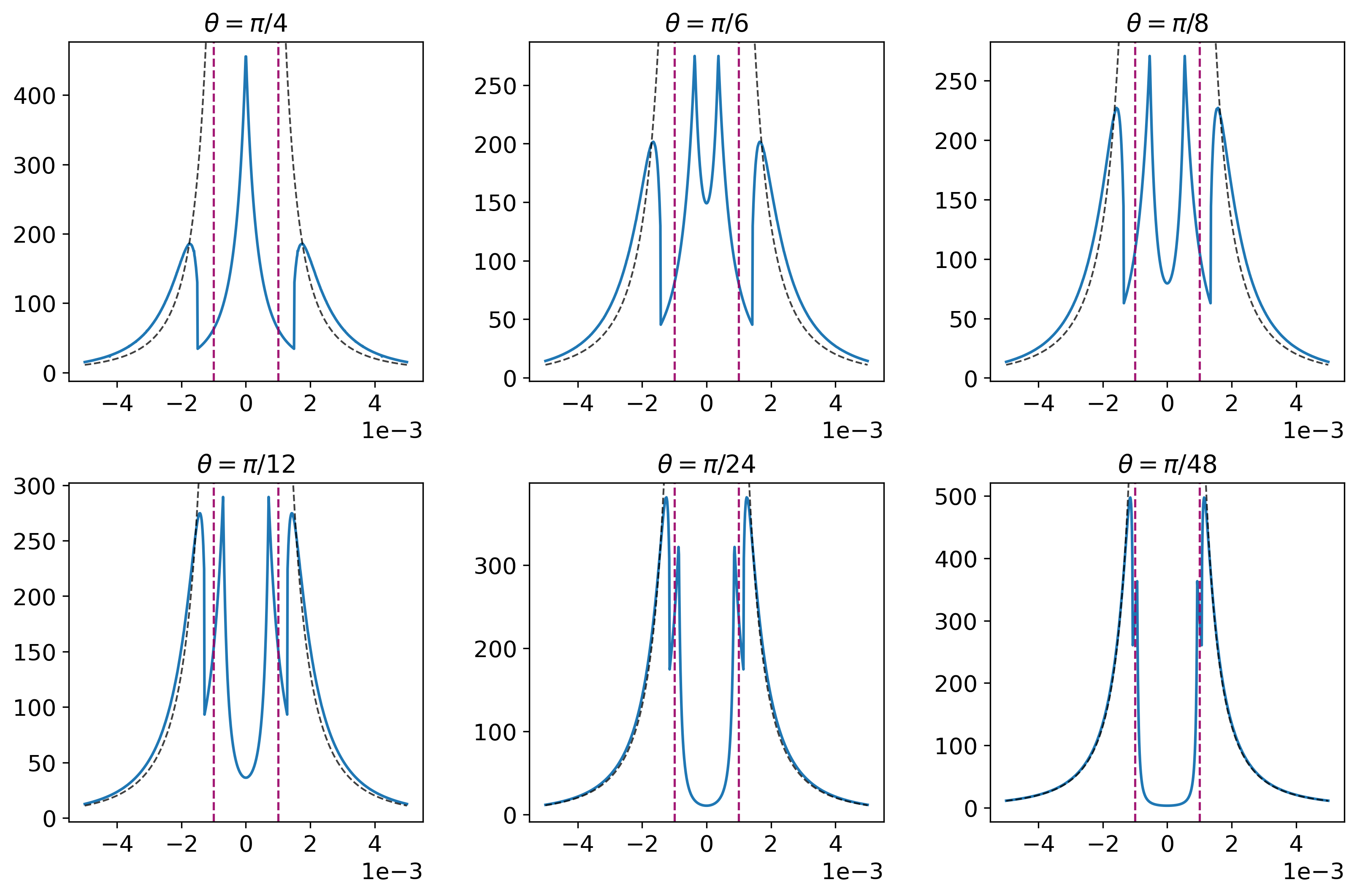}
    \caption{The convolution $\rho_\theta^\star$, defined in \cref{eq:rho_conv_definition}, for various values of $\theta$. The plots are generated for $\rho$ corresponding to the choice of $m = 1.7$, and $\delta = 10^{-3}$. While $m$ affects the steepness of the curves, $\delta$ has no effect on the nature of these curves. The vertical lines correspond to $\pm \delta$. Also shown (in gray) is the original distribution $\rho(t)$. Under the PPS hypothesis introduced in App. \ref{app:analytical}, coefficient distribution of Pauli terms belonging to $\mathcal P_k^{anti} \cap \sigma \mathcal P_k^{anti}$ is given by the truncation $\overline{(\rho_\theta)}$ of $\rho_\theta^\star$, defined as $\overline{(\rho_\theta)}(t) = \rho_\theta^\star(t)$ if $|t| \geq \delta$, 0 otherwise. As seen above, the effect of convolution is manifested as a reduction of density (depending on $\theta$) in the region between $\delta$ and 2$\delta$. We note that as $\theta \to 0$, $\rho_\theta^\star \to \rho$. Also, $\rho_\theta^\star$ is symmetric about $\theta = \pi/4$: $\rho_{\pi/4-\theta}^\star = \rho_{\pi/4+\theta}^\star$ and hence we have omitted plots for $\theta$ in the range $(\frac{\pi}{4}, \frac{\pi}{2})$.}
    \label{fig:rho conv various theta}
\end{figure}

It is important to tackle this problem for a fuller understanding of the dynamics of the distribution, as the approximation $\eta \ll 1$ is not always available. See \cref{fig:ac fracs rx_eq_random} for an example. To better understand the convolution $\rho_\theta^\star$, we first observe that it has the same asymptotic tail as $\rho(t)$. In fact, for any $t$ satisfying $|t| > 4\delta$, the approximation $\rho_\theta^\star(t) \approx \rho(t)$ is a good one, with the error in approximation getting smaller as we move along the tail of the distribution. We refer to \cref{fig:rho conv various theta} to understand the qualitative nature of these convolutions as $\theta$ varies.

\begin{figure}
    \centering
    \includegraphics[width=0.96\linewidth]{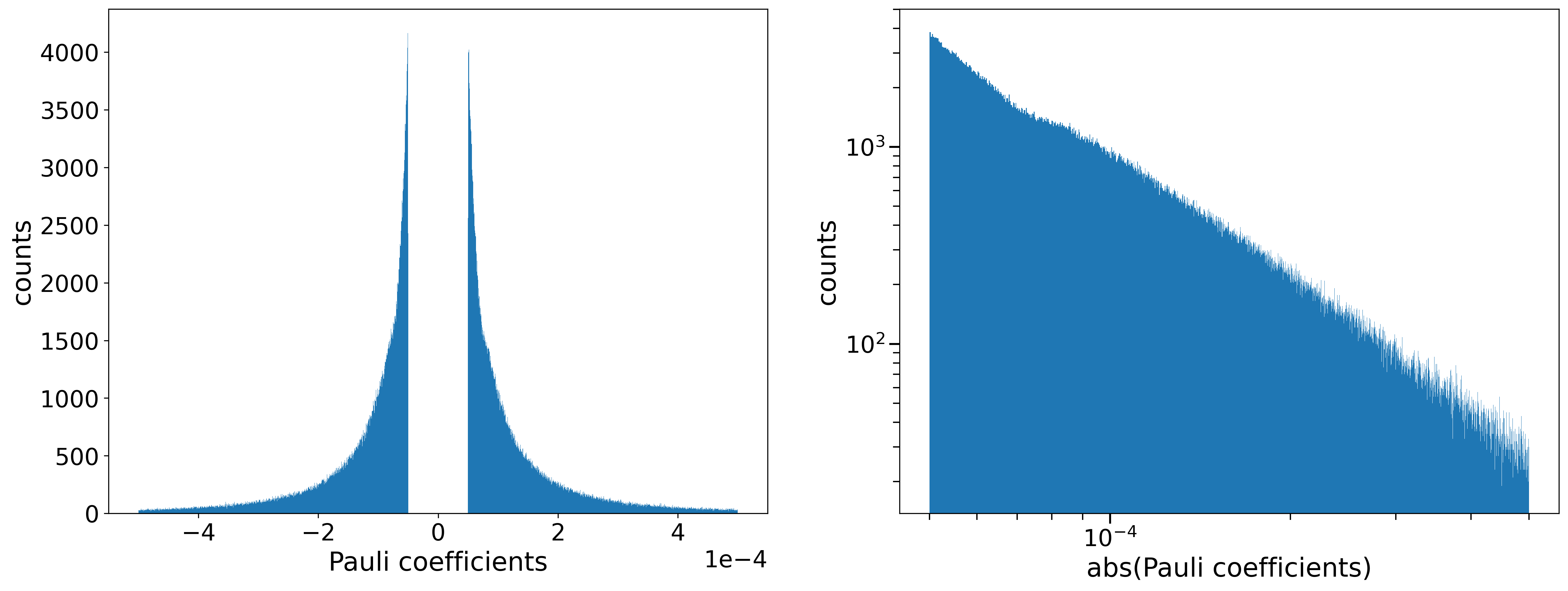}
    \caption{$\eta$-wiggle: These plots show the presence of a transient wiggle caused by an $\eta$-spike in the Pauli coefficient distribution $\rho$. The circuit we use is constructed with each $\theta_X$ sampled uniformly from $-[\pi/4,\pi/4]$ and $T= 30$. The simulation is performed using $\delta=5\times10^{-5}$. We notice a wiggle in the distribution of the absolute values of the Pauli coefficients between $\delta$ and $2\delta$, as explained in \cref{app:eta-spikes}. This snapshot of coefficients is taken after the application of 2036 of the 8130 gates. At the 2036th gate, there is an $\eta$-spike with $\eta \approx 0.368$ caused by a rotation gate with angle approximately $55.13^\circ$. However, this wiggle disappears quickly within the next few gate applications.}
    \label{fig:eta wiggle random rx}
\end{figure}

The plots in \cref{fig:rho conv various theta} and the sparseness of $\eta$-spikes in \cref{fig:ac fracs rx_eq_random} suggest the following explanation as to why the power-law distribution is stable under perturbation under some mild conditions: The truncation $\overline\rho_\theta$ of $\rho_\theta^\star$, which modifies $\rho$ as specified by \cref{eq: density_change_full}, shares the same tail of $\rho$, but diverges from $\rho$ in the region $\delta \leq |t| \leq 2\delta$, where it has a wiggle. This wiggle gets scaled by $\eta$ and is added to $\rho_{k+1}$; see \cref{fig:eta wiggle random rx}. However, since these perturbations occur close to the edge of the $\delta$-chasm, the wiggle quickly disappears (by branching followed by truncation) after the application of the next few gates. (Actually, a part of the wiggle lives on in the commutative part of the distribution. But after each subsequent gate application, this part gets diminished exponentially due to scrambling---as commuting Pauli terms for different gates are likely to be independent.)  

In the above argument, we have implicitly assumed that there is a gap to the next $\eta$-spike, and furthermore that the rotation angles of the gates that immediately follow are not too small (modulo $\pi/2$). We next discuss what may happen if these assumptions fail.  

\subsection{Small angles and persistent deviations}\label{app:large_deviations}

In this section we discuss scenarios where at some point in the circuit execution $\eta$-spikes occur in a consecutive and persistent manner. In such a situation, the distribution $\rho$ will be dominated by $\rho^\star_\theta$ and its higher convolutions ($(\rho_\theta^\star)^\star_\theta$, etc.), and this could alter the distribution $\rho$ in such a way that it does not recover its power-law shape. 

Even if the $\eta$-spikes are not consecutive, there can be cumulative effects seen over the course of many gate applications. The speed with which the wiggles produced by $\eta$-spikes get diminished (and eventually disappear) depends on the sequence of rotation angles of the subsequent gates. Since existing coefficients either remain the same (commutative case) or get multiplied by $\cos(\theta_j)$, smaller angles mean that the wiggles persist longer, thereby distorting the distribution via the compounding of self-convolutions. 

\begin{figure}
    \centering
    \includegraphics[width=0.96\linewidth]{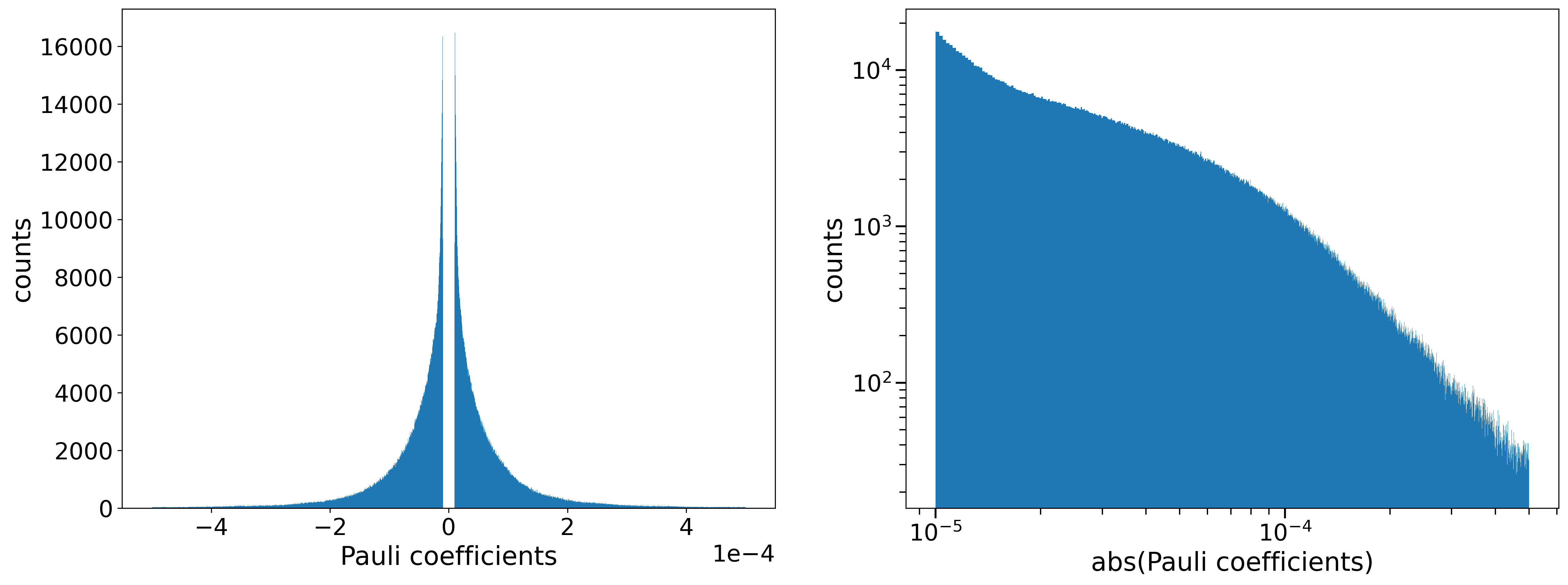}
    \caption{Plots corresponding to an experimental set up that exhibits visual deviations from a power-law distribution. Unlike in the rest of this work we choose the two qubit rotation angles to be non-Clifford. We set $\theta_{ZZ} = -\pi/36$, $\theta_X = \pi/12$, $T = 30$, and perform the simulation using $\delta = 10^{-5}$. The distributions shown are captured upon simulating all the gates in the circuit. Since the two-qubit rotations are chosen to be non-Clifford, they too contribute to $\eta$-spikes thus increasing their frequency. Because of the small angles involved, the effect of the convolution term in \cref{eq: density_change_full} persist longer, and this eventually leads to a distribution as shown above. The tail of the distribution, which contributes a majority to observable norm-squared, is still linear. We note here that the coefficient distribution for the $11\times11$ 2D Ising model circuit from \cite{PRXQuantum.6.020302} that we considered in \cref{subsec: N_max_in_practice} closely resembles the distribution above. Thus such a distorted distribution seems to arise intrinsically from the dynamics of \cref{eq:branching-and-merging} and furthermore appears to be stable.}
    \label{fig:deviation_from_power_law}
\end{figure}

We have in fact observed such instances of large deviation from the power law for modifications of the kicked Ising model IBM circuit where in all rotation angles, including the RZZ-angles, are small and correlated. See \cref{fig:deviation_from_power_law}. Even in such cases, we notice that the tail of the distribution follows a power law (with $x_{min}= l\delta$, for some $l$ independent of gate $k$). The estimates and analysis from \cref{subsec:power_law_bounds} carry over with only minor modifications to such cases.

\section{Challenges in estimating the power-law exponent $m$}
\label{app:m_estimation_challenges} 

\begin{figure}
    \centering
    \includegraphics[width=0.96\linewidth]{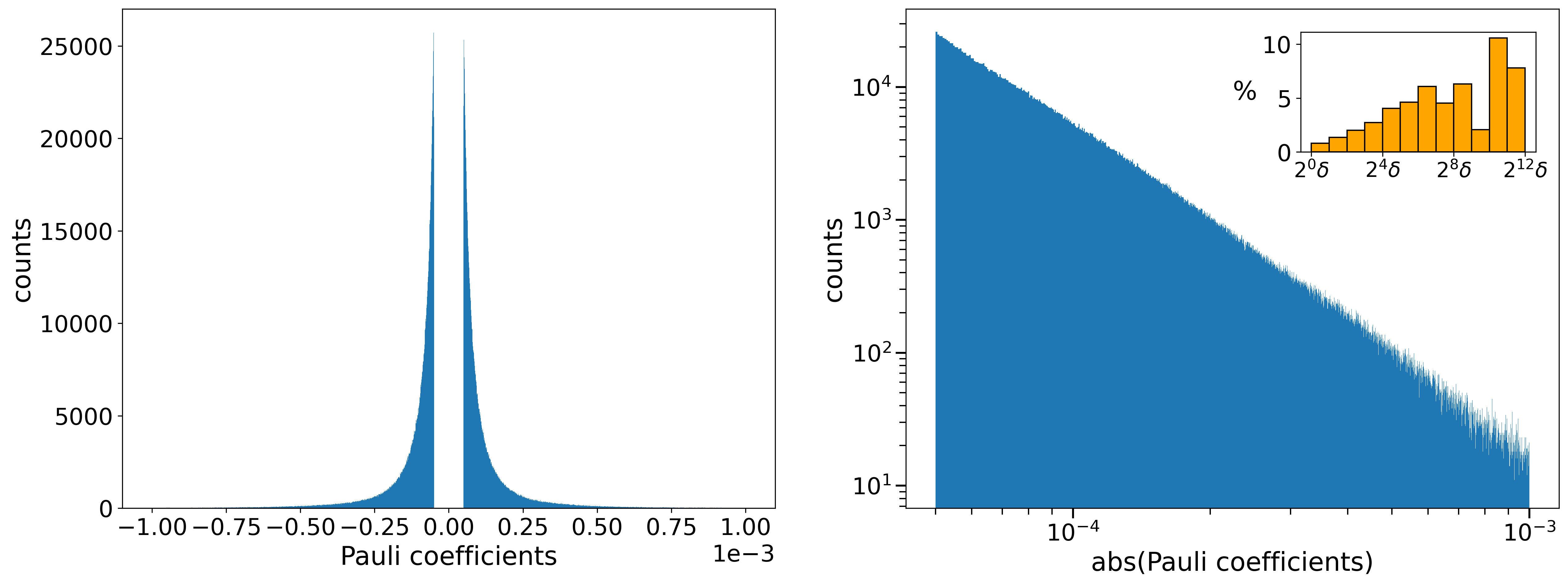}
    \caption{Distribution of Pauli coefficients for fixed $\theta_X$ equal to $\pi/6$, $T = 35$, and $\delta = 5\times10^{-5}$. The coefficients are captured at the end of the simulation. Inset (right): percentage contribution to norm-squared of the evolved observable, $\|O_k\|^2$ from coefficients $c_P$ belonging to intervals $2^j\delta \leq |c_P| < 2^{j+1}\delta$, for $j = 0, 1, 2, \ldots$. We note that there is a single coefficient that lies beyond $2^{13}\delta$ that contributes to around $47\%$ of the norm, which is not shown. We see that the tail of the distribution dominates $\|O_k\|^2$.}
    \label{fig: coeff distribution pi_by_6}
\end{figure}

We point out some challenges in estimating the power-law exponent $m$ for the coefficient distribution. First, we notice that when using linear regression on log-log plots, starting at $\delta$ leads to an underestimation of $m$, due to a subtle reduction in density in the interval between $\delta$ and $2\delta$. We observe this deviation even when the coefficient distribution visually resembles a power-law shape as in \cref{fig: coeff distribution pi_by_6}. We suspect the cumulative effects caused by $\eta$-spikes as the reason behind this phenomenon, as we explained in \ref{app:large_deviations} where we provided visually apparent examples of this effect.

\begin{figure}
    \centering
    \includegraphics[width=0.6\linewidth]{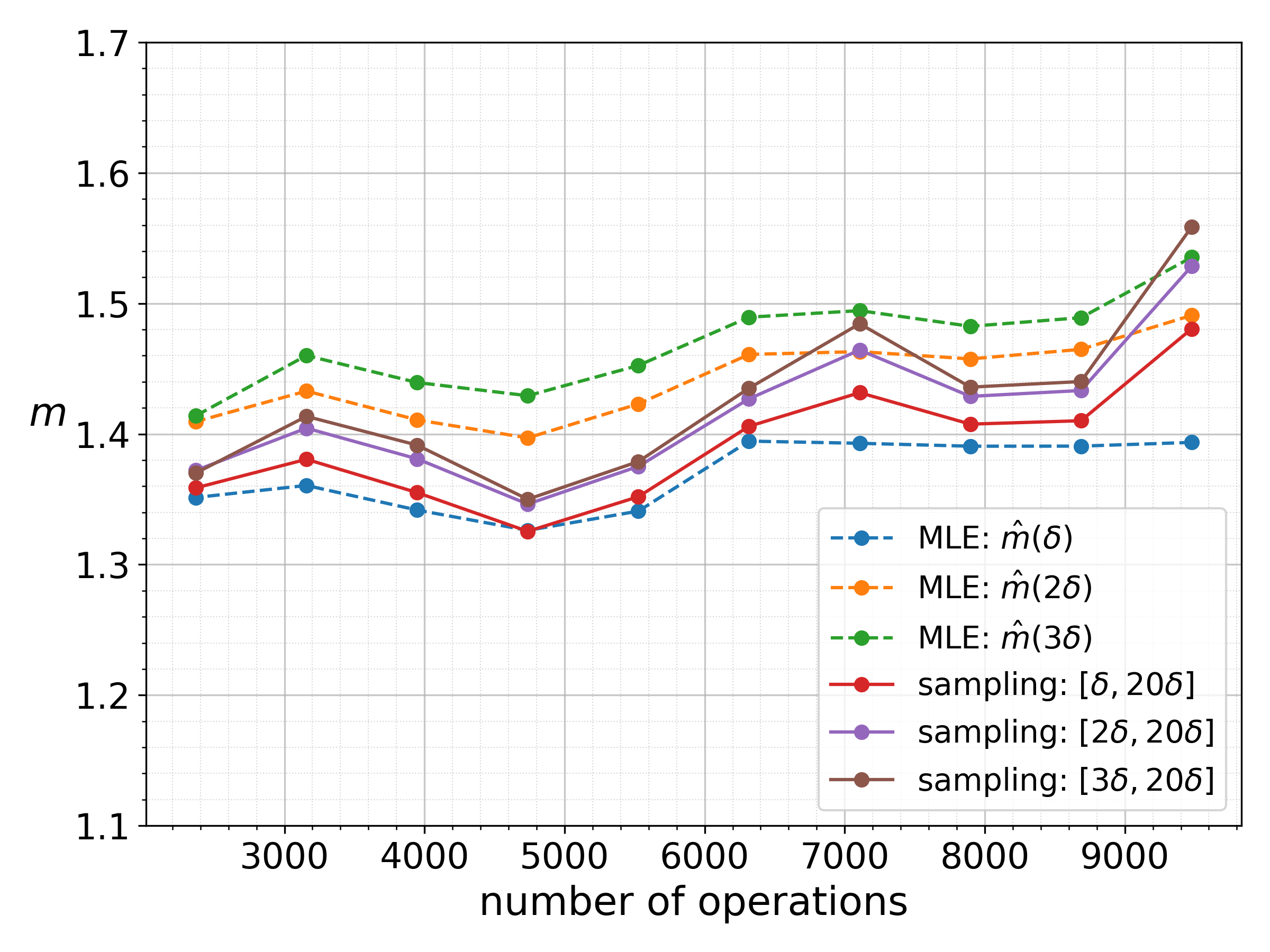}
    \caption{Plots showing variation of the power-law exponent $m$ within the circuit for the same circuit set up as in \cref{fig: coeff distribution pi_by_6}, and for different estimation techniques. For the sampling method, similar to \cref{fig:pauli-growth-and-varying-m}, we use bins of size $\delta/16$ centered at 144 uniformly spaced points $x$ between $l\delta$ and $20\delta$, for each $l = 1, 2$, and $3$, and use linear regression to find $m+1$ as the slope of log(density) vs $\log(x)$. Alternatively, one can use maximum likelihood estimators for $m$, $\hat m(l\delta)$, given by variants of \cref{eq:MLE}, with $x_{min} = l\delta$  instead of $\delta$. This amounts to ignoring all terms less than $l\delta$ (but keeping all of the tail).} 
    \label{fig:variation_of_m_with_mle}
\end{figure}

A more accurate approach to estimating $m$ is to use the maximum likelihood estimate (MLE), given by \begin{equation} 
\label{eq:MLE} 
\hat m_k = \frac{N_k}{\sum_{P} \log(|c_P^{(k)}|/\delta)}.
\end{equation} 
This also leads to a similar underestimation of $m$. See \cref{fig:variation_of_m_with_mle}.  Since a key motivation for estimating the distribution parameter is to have accurate estimates for $N_{\mathrm{max}}$ via \cref{eq: N_max bound}, perhaps another way to estimate $m$ would be a modification of the linear regression method, with the `continuous part' of the tail weighted more---with weights proportional to the contribution to the observable norm-squared, as in \cref{fig: coeff distribution pi_by_6}. (We did not attempt to implement this method.) 

Taking into account the aforementioned decrease in density in the leftmost part of the (absolute) distribution, we can parametrically estimate $m$, starting from $l\delta$, where $l = 1, 2, 3, \dots$ ($l$ can also be any real multiple of $\delta$). Likewise, one can consider maximum likelihood estimators $\hat m_k(l\delta)$, with $x_{min}$ set to $l\delta$; i.e., we replace $\delta$ with $l\delta$ in \cref{eq:MLE} and consider only those $P$ that satisfy $|c_P^{(k)}| \geq l\delta$. These estimates are shown in \cref{fig:variation_of_m_with_mle}. We did not explore the question of finding the optimal $l$ (or $x_{min}$) for the underlying distribution of Pauli coefficients $\{c_P^{(k)} \mid P \in \mathcal P_k\}$.  

We conclude by noting that the discrepancy between various estimates in \cref{fig:variation_of_m_with_mle} points to  statistical difficulties in estimating $m$ as well as to subtle deviations from the hypothesized power-law distribution.

\end{document}